\documentclass[journal,transmag,12pt,onecolumn,double]{IEEEtranTCOM}
\usepackage{inputenc}
\usepackage{amsmath}
\usepackage{amsfonts}
\usepackage{amssymb}
\usepackage{amsthm}
\usepackage{multirow}
\usepackage{tabulary}
\usepackage{booktabs}
\usepackage{graphicx}
\usepackage{bigstrut}
\usepackage{multirow}
\usepackage{tabulary}
\usepackage{filecontents}
\usepackage[square,comma,numbers,sort&compress]{natbib}
\usepackage{flafter}
\usepackage{lscape}
\usepackage{amssymb}
\usepackage{color}
\DeclareGraphicsExtensions{.pdf,.png,.gif,.jpg,.eps}
\usepackage{afterpage}
\usepackage{float}
\usepackage{rotating}
\usepackage{amsthm}
\usepackage{slashbox}
\usepackage{multirow}

\usepackage{multirow}
\usepackage{cases}
\usepackage{bm}
\usepackage{mathtools}
\usepackage{subfig}

\usepackage{setspace}
\newcolumntype{C}[1]{>{\centering\arraybackslash$}p{#1}<{$}}
\usepackage{tikz}
\usetikzlibrary{matrix}

\usepackage{fixltx2e}

\newcommand{\rotsim}{\mathbin{\rotatebox[origin=c]{30}{$\sim$}}}

\usepackage{hyperref}

\newcommand{\matfigwidth}{300pt}   

\usepackage{xfrac}
\DeclareDocumentCommand{\newfaktor}{s m O{0.5} m O{-0.5}}{
  \setbox0=\hbox{\ensuremath{#2}}
  \setbox1=\hbox{\ensuremath{\diagup}}
  \setbox2=\hbox{\ensuremath{#4}}
  \raisebox{#3\ht1}{\usebox0}
  \mkern-5mu\ifthenelse{\equal{#1}{\BooleanTrue}}
    {\diagup}
    {\rotatebox{-44}{\rule[#5\ht2]{0.4pt}{-#5\ht2+#3\ht0+\ht0}}}
  \mkern-4mu%
  \raisebox{#5\ht2}{\usebox2}
}

 \DeclareMathOperator*{\E}{\mathbb{E}}

\begin{document}

\newcounter{tempequationcounter}

 \newdimen\origiwspc%
  \newdimen\origiwstr%
  \origiwspc=\fontdimen2\font
  \origiwstr=\fontdimen2\font

\allowdisplaybreaks

\title{\LARGE Differential Modulation for Asynchronous Two-Way-Relay Systems over Frequency-Selective Fading Channels}

\author{
Ahmad~Salim,~\IEEEmembership{Student~Member,~IEEE,}
        and~Tolga~M.~Duman,~\IEEEmembership{Fellow,~IEEE}
\thanks{Ahmad Salim is with the School of Electrical, Computer and Energy Engineering (ECEE) of Arizona State University (ASU), Tempe, AZ 85287-5706, USA (E-mail: assalim@asu.edu).

Tolga M. Duman is with the Department of Electrical and Electronics Engineering (EEE), Bilkent University, Bilkent, Ankara, 06800, Turkey, and is an adjunct faculty with the School of ECEE of ASU.

This work was supported by the National Science Foundation under the grants NSF-CCF 1117174 and NSF-ECCS 1102357 and by the European Commission under the grant MC-CIG PCIG12-GA-2012-334213.
}	\vspace{-1.6cm}			
}
\maketitle
\begin{abstract}
In this paper, we propose two schemes for asynchronous multi-relay two-way relay (MR-TWR) systems in which neither the users nor the relays know the channel state information (CSI). In an MR-TWR system, two users exchange their messages with the help of $N_R$ relays. Most of the existing works on MR-TWR systems based on differential modulation assume perfect symbol-level synchronization between all communicating nodes. However, this assumption is not valid in many practical systems, which makes the design of differentially modulated schemes more challenging. Therefore, we design differential modulation schemes that can tolerate timing misalignment under frequency-selective fading. We investigate  the performance of the proposed schemes in terms of either probability of bit error or pairwise error probability. Through numerical examples, we show that the proposed schemes outperform existing competing solutions in the literature, especially for high signal-to-noise ratio (SNR) values.

\end{abstract}

\vspace{-.25cm}

\vspace{-.5cm}
\section{Introduction} \label{intro_Diffsmall}

Most of the existing schemes for TWR systems assume known CSI (see, e.g., \cite{Salim_2015,Duman_2015} and the references therein). Due to many reasons, such as the large overhead of channel estimation process or relatively rapid variations of the channel, perfect CSI is not always available. In such scenarios, using non-coherent modulation schemes such as differential phase shift keying (DPSK) that require no CSI knowledge is a practical solution. 

While there have been significant research efforts on using differential modulation (DM) for TWR systems, most, e.g. \cite{Song_2010}, assume symbol-level synchronization among all nodes. In practice, many reasons such as different propagation delays or different dispersive channels, lead to a timing misalignment between the arriving signals. Therefore, having a perfectly synchronized TWR system is very difficult which, in return, renders the design of differentially modulated schemes more challenging.
In the case of synchronous TWR systems, many schemes were proposed to address the absence of CSI, e.g. \cite{Cui_2009,Song_2010,Guan_2011,Zhu_2012}. However, little work has been conducted to tackle asynchronous communication scenarios. One scenario of particular interest is the use of asynchronous MR-TWR systems in which timing errors not only occur at users but at relays as well.

In \cite{Cui_2009}, the authors propose a DM scheme along with maximum likelihood (ML) detection and several suboptimal solutions for a number of relaying strategies when CSI is not available at any node. The authors further extend their results to the multi-antenna case based on differential unitary space-time modulation. A simple amplify-and-forward (AF) scheme is proposed in \cite{Song_2010} based on DM in which the self-interference term is estimated and removed prior to detection. The resulting bit error rate (BER) and the optimum power allocation strategies are also studied. In \cite{Debbah2010}, the authors propose a joint relay selection and AF scheme using DM. The scheme selects the relay that minimizes the maximum BER of the two sources. Ref. \cite{Guan_2011} proposes a DM scheme that uses $K$ parallel relays, for which a denoising function is derived to detect the sign change of the network coded symbol at each relay which is used later by the users for detection. The paper obtains a closed form expression for the BER for the single-relay case along with deriving a sub-optimal power allocation scheme. Furthermore, the authors derive lower and upper bounds on the BER for the multi-relay case.
A low complexity DPSK-based scheme is proposed in \cite{Zhu_2012} for physical-layer network coding to acquire the network coded symbol at the relay without requiring CSI knowledge. Compared to the schemes in \cite{Cui_2009,Guan_2011} which require more complexity, this scheme shows better performance at high SNRs. However, the detector is only derived for a binary alphabet.

A few proposals in the literature considered the design of distributed space time coding (DSTC) coupled with differential modulation for synchronous TWR systems, e.g., \cite{utkovski_2009_distributed,Huo_DDSTC_TWR_2012,alabed_2013_distributed}. The models in \cite{utkovski_2009_distributed,Huo_DDSTC_TWR_2012} assume two-phase transmission and the lack of a direct link between the two users. On the other hand, \cite{alabed_2013_distributed} assumes a three-phase transmission and that a direct link between the two users exist.

All the solutions discussed above have strict synchronization requirements for proper operation. Only few works considered asynchronous TWR systems where DM is used to mitigate CSI absence. For instance, \cite{Zhuo_diff_asynch_TWR_2013} proposes an interference cancellation scheme to reduce the interference from neighboring symbols caused by imperfect synchronization. Ref. \cite{qian2014asynchronous} extends the scheme in \cite{Zhuo_diff_asynch_TWR_2013} to dual-relay TWR systems.

While \cite{Zhuo_diff_asynch_TWR_2013,qian2014asynchronous} present important results, they are restricted to flat fading channels, and the delays that can be tolerated are only within the period of a symbol, which make them suitable neither for time-dispersive channels nor for systems experiencing large relative propagation delays. In this paper, we consider a more general frequency-selective fading channel and propose two schemes that can tolerate larger relative propagation delays compared to \cite{Zhuo_diff_asynch_TWR_2013}. Specifically, we first propose the joint blind-differential (JBD) detection scheme in which we first perform blind channel estimation to be able to remove the self-interference component, and then perform differential detection. We provide an approximate closed form expression for the BER for large SNR values. We then propose a scheme that is based on differential DSTC, referred to as JBD-DSTC, to fully harness the available diversity in the system. The JBD-DSTC scheme significantly reforms the JBD scheme in order to obtain an STC structure for the partner's message at each user. The pairwise error probability of this scheme along with the achievable diversity is also discussed.

The remainder of this paper is organized as follows. Section \ref{sec:system_model_Diffsmall} describes the system model. Section \ref{sec:JBD} details the transmission mechanism and receiver design for the proposed JBD scheme along with providing a closed form expression for the probability of error. Section \ref{sec:JBD-DSTC} presents the JBD-DSTC scheme and the relevant performance analysis in terms of the PEP. Section \ref{sec:performanceecval_Diffsmall} presents numerical results obtained to evaluate the performance of the proposed solutions. Finally, conclusions are drawn in Section \ref{sec:conclusions_Diffsmall}.

Notation: Unless stated otherwise, bold-capital letters refer to frequency-domain vectors, bold-lower case letters refer to time-domain vectors, capital letters refer to matrices or elements of frequency-domain vectors (depending on the context), and lower-case letters refer to scalars or elements of time-domain vectors. If used as a superscript, the symbols $T$, $*$ and $H$ refer to transpose, element-wise complex conjugate and Hermitian transpose (conjugate transpose), respectively. The notation ${\bm 0}_N$ and $0_{N\times N}$ refer to length-$N$ all-zero column vector and all-zero matrix, respectively. $F$ is the normalized discrete Fourier transform (DFT) matrix of size-$N$. The Inverse DFT (IDFT) matrix of size-$N$ is denoted by $F^H$. The subscript $ir$ refers to the channel from node $i$ to node $r$.

\vspace{-1mm}	
\section{System Model} \label{sec:system_model_Diffsmall}
\begin{figure} %
\vspace{-0.2in}
\centering\includegraphics[trim =30mm 90mm 12mm 139mm, clip, scale=0.6] {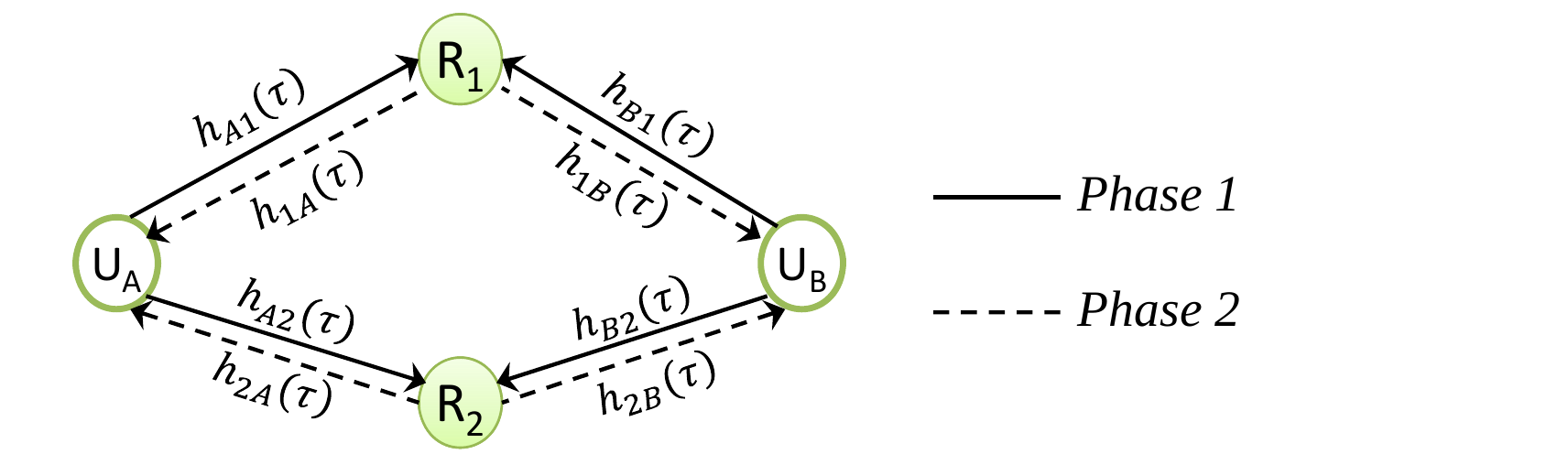}
\vspace{-0mm} \caption{The MR-TWR system model (for $N_R=2$).}
 \label{fig:DTWR_dsmall}
 \vspace{-0mm}
 \end{figure}

We consider a two-phase communication scheme using AF relaying (as shown in Fig. \ref{fig:DTWR_dsmall} for the case of two relays). The users exchange data by first simultaneously transmitting their messages to the relays during the multiple-access (MAC) phase. During the broadcast (BC) phase, each relay broadcasts an amplified version of its received signal which is a noisy summation of the users' messages.

Each user transmits $M$ blocks that comprise one frame. Prior to transmission, each block is modulated using orthogonal frequency division multiplexing (OFDM) with $N$ subcarriers. Each one of the resulting blocks is appended with a cyclic-prefix (CP). We model asynchrony by assuming different propagation delays. For proper CP design, user $U_i$, $i\in\{A,B\}$, requires the knowledge of the worst-case scenario propagation delays over the links connecting it to the relays, i.e., $d_{ir}$ (in multiples of the sampling time), $r\in\{1,2,\ldots,N_R\}$. Similarly, the $r$th relay, $r\in\{1,2,\ldots,N_R\}$, requires $d_{ri}$, $i\in\{A,B\}$.

The multipath fading channels from the users to the relays are modeled (in the equivalent low-pass signal domain) by the discrete channel impulse responses (CIRs) $h_{ir,l}$, $i\in\{A,B\}$, $r\in\{1,2,\ldots,N_R\}$, $l\in\{1,2,\ldots,L_{ir}\}$, where $L_{ir}$ represents the number of resolvable paths. Similarly, the channels from the relays to the users are modeled by $h_{ri,l}$.
The overall channel response over the $L_{ir}$ lags can be expressed as
${h}_{ir}(\tau) =  \sum_{l=1}^{L_{ir}}{h_{ir,l}\delta \left( \tau-\tau_{ir,l} \right) }$,
where $\tau$ is the lag index and $\tau_{ir,l}$ is the delay of the $l^{th}$ path normalized by the sampling period $T_S$.
We assume quasi-static frequency-selective fading in which $ h_{ir,l}$ remain constant for all the blocks over the same lag ($l$) and change independently across the different lags. We assume that $ h_{ir,l}$ is a circularly-symmetric complex Gaussian (CSCG) random variable (RV) with zero mean and variance of $\sigma_{ir,l}^2$. Also, the channel coefficients are independent across different links.
Further, we assume half-duplex operation at all nodes.

For the JBD scheme, we further assume that the channels on the same link are reciprocal, i.e., ${h}_{ir}(\tau)={h}_{ri}(\tau)$ $\forall i,r$. Also, the uplink and downlink propagation delays over the same link are assumed to be identical. 

\vspace{-0mm}
\section{The Joint Blind-Differential (JBD) Scheme} \label{sec:JBD}
In this scheme, each user uses $N$ parallel differential encoders each operating on a specific subcarrier. The data vector representing the frequency-domain message of the $i$th user, $i\in\{A,B\}$, during the $m$th block is denoted by ${  \bm X}_{i}^{(m)}$ where ${  \bm X}_{i}^{(m)}=\left[ {X}_{i,1}^{(m)} ,{X}_{i,2}^{(m)}, \ldots,{X}_{i,N}^{(m)}\right]^T$ and ${X}_{i,k}^{(m)} \in{\mathcal{A}_i}$ where $\mathcal{A}_i$ is a unit-energy, zero-mean, phase-shift keying (PSK) constellation set that is closed under multiplication, e.g., the set $\{\pm 1,\pm j\}$, to maintain the transmit power at a specific level. Using DM, the differentially encoded symbol over the $k^{th}$ subcarrier of the $m$th block can be expressed as
${S}_{i,k}^{(m)} = {X}_{i,k}^{(m)}  {S}_{i,k}^{(m-1)}$, $m\in\{2,3,\ldots,M\}$,
After performing IDFT, we obtain ${\bm s}_{i}^{(m)}=\left[ {s}_{i,1}^{(m)} ,{s}_{i,2}^{(m)}, \ldots,{s}_{i,N}^{(m)}\right]^T={\rm IDFT} ( {  \bm S}_{i}^{(m)})$.
The transmitted signal from the $i$th user during the $m$th block, $i\in\{A,B\}$, is given by:
\begin{equation}
{  \bm s}_{Tx,i}^{(m)} = \sqrt{P_{i}} \zeta_1\left(  {\bm s}_{i}^{(m)} \right)
\end{equation}
where ${  \bm s}_{Tx,i}^{(m)}=\left[ {s}_{Tx,i,1}^{(m)} ,{s}_{Tx,i,2}^{(m)}, \ldots,{s}_{Tx,i,N+N_{CP,1}}^{(m)}\right]^T$, $P_{i}$, $i\in\{A,B\}$, is the transmission power at the $i$th user and $\zeta_1(\cdot)$ corresponds to the operation of appending a length $N_{CP,1}$ CP to the vector in its argument at each user prior to the first phase of transmission. The length of this CP is selected to satisfy $N_{CP,1} \geq \max_{i,r}\{L_{ir} + d_{ir} \}$, $i\in\{A,B\}$, $r\in\{1,2\}$.

\subsection{Relay Processing}\label{subsec:JBD_RelayProcessing}
Having appended a CP of the proper length at each user, the received signal corresponding to the $m$th block at the $r$th relay after removing the CP is given by
\begin{equation} \label{eqn:yj}
\notag \begin{array}{rll}
 {  \bm y}_{r}^{(m)}
\hspace{-1mm} =\sqrt{P_A}  H_{tl,Ar} \Psi_{d_{Ar}} {  \bm s}_{A}^{(m)}\hspace{-1mm}+\hspace{-1mm}\sqrt{P_B}  H_{tl,Br} \Psi_{d_{Br}} {  \bm s}_{B}^{(m)}\hspace{-1mm}+\hspace{-.5mm}{  \bm n}_{r}^{(m)},
 \end{array}
\end{equation}
where $H_{tl,ir}$ is the time-lag channel matrix corresponding to the channel over the link $ir$ and ${\bm n}_{r}^{(m)}$ represents length-$N$ noise vector at the $r$th relay during the $m$th block whose entries are independent and identically distributed (i.i.d.) CSCG random variables (RVs) with zero mean and variance of $\sigma_{ r}^2$. $\Psi_{d_{ir}} $, $i\in\{A,B\}$, $r\in\{1,2,\ldots,N_R\}$, is a circulant matrix of size $N \times N$ whose first column is given by the $N \times 1$ vector $\psi_{d_{ir}}=[ {\bm 0}^T_{d_{ir}},1, {\bm 0}^T_{N-d_{ir}-1}]^T$. Using the matrix $\Psi_{d_{ir}}$  mimics the circular shift caused by having a propagation delay of $d_{ir}$ samples.

To simplify blind channel estimation at the end user, $R_r$ performs conjugation and time-reversal operations to obtain ${  \bm s}_{r}^{(m)} =  \eta\left( {\bm y}_{r}^{(m)*} \right)$ where $\eta(\cdot)$ is the time-reversal operator. For ${\bm x}=\left[ {x}_{1} ,{x}_{2}, \ldots,{x}_{N}\right]^T$, $\eta(\cdot)$ is defined element-wise as $\eta(x_n)\triangleq x_{N-n+2} $, $n=1,\ldots,N$ and $x_{N+1}\triangleq x_{1}$.
 The conjugation and reversal in the time-domain will have a conjugation effect in the frequency-domain after taking DFT at the end user.

After processing the mixture of signals, $R_r$ appends a CP for the second phase of transmission of length $N_{CP,2}$ that satisfies $N_{CP,2} \geq \max_{r,i}\{L_{ri} + d_{ri} \}$, $r\in\{1,2,\ldots,N_R\}$, $i\in\{A,B\}$. The $r$th relay transmitted signal is given by:
\begin{equation}
{  \bm s}_{Tx,r}^{(m)} = \sqrt{P_r G_{r}} \zeta_2\left(  {\bm s}_{r}^{(m)} \right) , \quad r\in\{1,2\}
\end{equation}
where ${  \bm s}_{Tx,r}^{(m)}=\left[ {s}_{Tx,r,1}^{(m)} ,{s}_{Tx,r,2}^{(m)}, \ldots,{s}_{Tx,r,N+N_{CP,2}}^{(m)}\right]^T$, $P_{r}$ and $G_r$ are the transmission power and the scaling factor at the $r$th relay, respectively, and $\zeta_2(\cdot)$ corresponds to the operation of appending a length $N_{CP,2}$ CP to the vector in its argument.

\subsection{Detection at the End-User}\label{subsec:JBD_Detection}
Due to symmetry, we only describe detection at user $B$. After removing the CP that was added at the relays, the received $N$-sample OFDM blocks can be written as
\begin{equation} \label{eqn:yB}
\notag \begin{array}{lll}
 {\bm y}_{B}^{(m)}
 =& \displaystyle \sum_{r=1}^{N_R}{ \sqrt{P_A P_r G_r}  H_{tl,rB} \Psi_{d_{rB}} \eta\left(  H_{tl,Ar}^{*} \Psi_{d_{Ar}}^* {\bm s}_{A}^{(m)*}\right) }\\
& \hspace{0mm} + \displaystyle \sum_{r=1}^{N_R}{\sqrt{P_B P_r G_r}   H_{tl,rB} \Psi_{d_{rB}}\eta\left( H_{tl,Br}^{*} \Psi_{d_{Br}}^* {\bm s}_{B}^{(m)*}\right)} +{  \bm v}_{B}^{(m)},
 \end{array}
\end{equation}
where ${\bm v}_{B}^{(m)}$ represents length-$N$ effective noise vector at user $B$ during the $m$th block which encompasses the relays’ ampliﬁed noise as well. The entries of ${\bm v}_{B}^{(m)}$ are i.i.d. CSCG RVs with zero mean and variance of $\sigma_{B,eff}^2=\sigma_{ B}^2 + \sum_{r=1}^{N_R}{G_{r} P_{r} \left| \left[  H_{df,rB}\right]_{k,k} \right|^2 \sigma_{r}^2}$ where $\sigma_{ B}^2$ is the variance of the original noise terms at user B.

Let ${  \bm V}_{B}^{(m)}=F {  \bm v}_{B}^{(m)}$, $P_{ir}=P_i P_r G_r$ and assume that $d_{ri}=d_{ir}$, $r\in\{1,2,\ldots,N_R\}$, $i\in\{A,B\}$. After performing DFT and noting that $F \eta\left( {\bm x^*}\right)=\left( F {\bm x}\right)^*$,

the received signal on the $k^{th}$ subcarrier of the $m$th block simplifies to\footnote{Refer to Appendix \ref{sec:Diff_appendixA} for details.} $Y_{B,k}^{(m)} =  \mu_{k} { { S}_{B,k}^{(m)}}^* +
 \nu_{k}  { { S}_{A,k}^{(m)}}^* +{V}_{B,k}^{(m)}$
where

\begin{eqnarray}
\notag \nu_{k} = \displaystyle \sum_{r=1}^{N_R}{\sqrt{P_{Ar}} \left[  H_{df,rB}\right]_{k,k} \left[H_{df,Ar}^* \right]_{k,k} e^{-j\frac{ 2\pi \left( k-1 \right) \left( d_{rB}-d_{Ar} \right)}{N}} } ,
\end{eqnarray}
$\mu_{k} =\sum_{r=1}^{N_R}{\sqrt{P_{Br}}\left|[ H_{df,Br}]_{k,k}\right|^2 }$, ${ V}_{B,k}^{(m)}$ is the $k^{th}$ element of ${\bm V}_{B}^{(m)}$ and $H_{df,ir} = F H_{tl,ir} F^H$ denotes the Doppler-frequency channel matrix (also called the subcarrier coupling matrix) over the link $ir$ which is a diagonal matrix in our case of quasi-static fading. 

The results of \cite{Song_2010} are adopted to estimate the parameter $\mu_{k}$ in order to remove the self-interference term.
Defining, $\widetilde{Y}_{B,k}^{(m)}= { { X}_{B,k}^{(m)}}^* Y_{B,k}^{(m-1)} - Y_{B,k}^{(m)}$, we can write
\begin{equation} \label{eqn:ytildeBkm}
\widetilde Y_{B,k}^{(m)} = \nu_{k} { { S}_{A,k}^{(m-1)}}^* \left( {{X}_{B,k}^{(m)}}^* - {{X}_{A,k}^{(m)}  }^*\right) +\widetilde{V}_{B,k}^{(m)}, \quad  m=2,\ldots,M,
\end{equation}
where $\widetilde{V}_{B,k}^{(m)}={{X}_{B,k}^{(m)}}^* { V}_{B,k}^{(m-1)} - { V}_{B,k}^{(m)}$. At high SNR, we can approximate
$\mbox{${\widetilde Y}^{(m)}_{B,k}$}^* \widetilde Y_{B,k}^{(m)}$ as
\begin{equation}  \label{eqn:ytcyt}
 \mbox{${\widetilde Y}^{(m)}_{B,k}$}^* \widetilde Y_{B,k}^{(m)} \approx \left| \nu_{k}   \right|^2 \left| { {S}_{A,k}^{(m-1)}} \right|^2 \left| {{X}_{B,k}^{(m)}} - {{X}_{A,k}^{(m)}  }\right|^2 , \quad  m=2,\ldots,M.
\end{equation}
Taking the expected value of \eqref{eqn:ytcyt} over the constellation points of ${S}_{A,k}^{(m-1)}$, ${X}_{A,k}^{(m)} $ and ${X}_{B,k}^{(m)}$, we note that for the RHS, it is the same for all $m$ and $k$ since the constellation sets $\mathcal{A}_i$, $i\in\{A,B\}$ are the same for all blocks and subcarriers. We also note that ${S}_{A,k}^{(m-1)}$ is independent from both ${X}_{B,k}^{(m)}$ and ${X}_{A,k}^{(m)}$.
For a sufficiently large $M$, we can approximate the ensemble average of $\mbox{${\widetilde Y}^{(m)}_{B,k}$}^* \widetilde Y_{B,k}^{(m)}$ by its time average. Therefore, we can obtain an estimate of $\left| \nu_{k}   \right|$, denoted by ${\left| \widehat{\nu}_{k}   \right|}$, as
\begin{equation}  \label{eqn:v1pv2}
{\left| \widehat{\nu}_{k}   \right|}^2
  \approx  \frac{\displaystyle \sum_{m=2}^{M}    {\left|\widetilde Y_{B,k}^{(m)}\right|}^2   }{(M-1)\E \left[ \left| { { S}_{A,k}^{(m-1)}} \right|^2  \right] \E \left[  \left| {{X}_{B,k}^{(m)}} - {{X}_{A,k}^{(m)}  }\right|^2\right]   }  ,
\end{equation}
where $\E \left[ \left| { { S}_{A,k}^{(m-1)}} \right|^2  \right]=1$ and $\E \left[  \left| {{X}_{B,k}^{(m)}} - {{X}_{A,k}^{(m)}  }\right|^2\right]$ can be calculated easily since the corresponding set defined by $\mathcal{K}=\left\{|b-a|^2 \; |\; b\in \mathcal{A}_B, a\in \mathcal{A}_A\right\}$ is finite. For instance, if $\mathcal{A}_i=\{1,-1\}$, $i\in\{A,B\}$, then $\mathcal{K}=\left\{0,4\right\}$ and $\E \left[  \left| {{X}_{B,k}^{(m)}} - {{X}_{A,k}^{(m)}  }\right|^2\right]=2$.
Let $\bm Y_{B,k}=\left[ {Y}_{B,k}^{(1)},{Y}_{B,k}^{(2)},\ldots, {Y}_{B,k}^{(M)} \right]^T$.
If $M$ is sufficiently large, we can approximate $\bm Y_{B,k}^{H} \bm Y_{B,k}$ as
\begin{equation}
\bm Y_{B,k}^{H} \bm Y_{B,k} \approx
M\left(
 \mu_{k}^2
+\left| \nu_{k}  \right|^2 + \sigma_{V_B}^2
 \right).
\end{equation}
At high SNR, we can write
\begin{equation}
 \mu_{k}^2  +\left| \nu_{k}  \right|^2 \approx   \frac{ \bm Y_{B,k}^{H} \bm Y_{B,k}}{M}.
\end{equation}
Therefore, we can estimate $\mu_{k}$ as
\begin{equation}
\widehat{\mu}_{k} \approx \sqrt{ \left( \frac{ \bm Y_{B,k}^{H} \bm Y_{B,k}}{M} -  {\left| \widehat{\nu}_{k}   \right|}^2 \right)  \mathrm{U} \left(\frac{ \bm Y_{B,k}^{H} \bm Y_{B,k}}{M} - {\left| \widehat{\nu}_{k}   \right|}^2  \right)  },
\end{equation}
where $\mathrm{U} \left(. \right) $ is the Heaviside unit step function. 
Now, we can remove the estimated self-interference term, namely $\widehat{\mu}_{k} { { S}_{B,k}^{(m)}}^*$ to obtain
 \begin{equation} \label{eqn:YAB} 
\begin{array}{lll}
Y_{AB,k}^{(m)}
 &\triangleq & Y_{B,k}^{(m)}- \widehat{\mu}_{k} { { S}_{B,k}^{(m)}}^* \\
  & \approx & \nu_{k} { { S}_{A,k}^{(m)}}^* +{V}_{B,k}^{(m)}, \quad  m=1,\ldots,M. \\
 \end{array}
\end{equation}

We can further express $Y_{AB,k}^{(m)}$ as
\begin{equation} \label{eqn:yBkm_1_small_noSB}
Y_{AB,k}^{(m)} \approx  { {X}_{A,k}^{(m)}}^* Y_{AB,k}^{(m-1)}
+\left(  {V}_{B,k}^{(m)}- { { X}_{A,k}^{(m)}}^*{V}_{B,k}^{(m-1)}\right), \quad  m=2,\ldots,M.
\end{equation}
Therefore, we write the following symbol-by-symbol MLD rule to recover ${X}_{A,k}^{(m)}$ at user B
  \begin{align}
      \widehat{X}_{A,k}^{(m)} &=\arg\,\underset{X\in{\mathcal{A}_A}}{\min}\left| Y_{AB,k}^{(m)} - { X }^* Y_{AB,k}^{(m-1)} \right|^2\\
      &=\arg\,\underset{X\in{\mathcal{A}_A}}{\max}\,\text{Re}\left\{ Y_{AB,k}^{(m)} {Y_{AB,k}^{(m-1)}}^* X \right\} , \quad  m=2,\ldots,M.
   \label{Eq:mldafk}
  \end{align}
We remark that better performance can be attained if multiple-symbol differential detection, as in \cite{Divsalar_1990}, is used. However, the detection complexity will be greater. 

\subsection{Performance Analysis} \label{Pe_JBD}
In this section we provide an approximate closed form expression for the probability of error of the JBD scheme by using results from the frequency-flat, Rayleigh-faded, single-way relay systems in \cite{Hasna_2003,Song_2010}.

Assume that instead of using $G_{r}$ to normalize the power at the $r$th relay in time domain, we use $G_{r,k}$ to normalize the power of the $k$th subcarrier in frequency domain. Note that $G_{r,k}$ can be estimated for large $M$ as $G_{r,k}\approx \frac{M}{||\bm Y_{r,k}||^2}$ without any CSI knowledge at the relay where $\bm Y_{r,k}=\left[ {Y}_{r,k}^{(1)},{Y}_{r,k}^{(2)},\ldots, {Y}_{r,k}^{(M)} \right]$ and ${  \bm Y}_{r}^{(m)}=\left[ {Y}_{r,1}^{(m)} ,{Y}_{r,2}^{(m)}, \ldots,{Y}_{r,N}^{(m)}\right]^T={\rm DFT} ( {  \bm y}_{r}^{(m)})$. By modeling the JBD system by an equivalent coherent receiver with treating $\nu_{k}$ as a known channel gain and $\left(  {V}_{B,k}^{(m)}- { { X}_{A,k}^{(m)}}^*{V}_{B,k}^{(m-1)}\right)$ as the equivalent noise term, we can approximate the effective SNR over the $k$th subcarrier at user B as
\begin{IEEEeqnarray}{rCl} \label{eqn:SNR_B}
\gamma_{B,k} &\approx & \frac{\left|\nu_{k}\right|^2 }{2 \text{Var}\left[ {V}_{B,k}^{(m)} \right]}\\
\label{eqn:SNR_B_2}
&=& \frac{ P_{A}\displaystyle \sum_{r=1}^{N_R}{ P_r \left| q_{rB,k} \right|^2 \left| q_{Ar,k} \right|^2
+  P_A\displaystyle \sum_{i=1}^{N_R}\sum_{j=1,j\neq i}^{N_R}\sqrt{P_{i}P_{j}G_{i,k}G_{j,k}} q_{iB,k} q_{Ai,k}^* q_{jB,k}^* q_{Aj,k}
}}{2 \left( \sigma_{ B}^2 + \sum_{r=1}^{N_R}{G_{r,k} P_{r} \left| q_{rB,k} \right|^2 \sigma_{r}^2} \right)},
\end{IEEEeqnarray}
where $q_{ij,k}=\left[  H_{df,ij}\right]_{k,k}$ and $\text{Var}[\cdot]$ is the variance operator.

Since $\gamma_{B,k}$ in \eqref{eqn:SNR_B} is a complicated function of $2N_R$ Rayleigh-distributed RVs, finding its statistics (PDF, CDF, etc.) is difficult, and hence deriving the probability of error is intractable. However, an important result in \cite{Hasna_2003} for a special choice of the scaling factor simplifies the analysis as it results in expressing the effective SNR in terms of the harmonic mean of the instantaneous SNR of the two hops, which in turn simplifies the calculations. The adopted scaling factor normalizes the power of the $k$th subcarrier as $G_{r,k}=\left( P_A \left| \left[ H_{df,Ar} \right]_{k,k}\right|^2  + P_B \left| \left[ H_{df,Br} \right]_{k,k}\right|^2 + \sigma_r^2 \right)^{-1}$. At this point, we adopt this scaling factor to make the analysis tractable for the JBD scheme.

Assume that $\sigma_i^2=\sigma_r^2=\sigma^2$ $\forall$ $i\in\{A,B\}$, $r\in\{1,2,\ldots,N_R\}$ and let $\gamma_{1}=\frac{P_A}{\sigma^2}$ and $\gamma_{2}=\frac{  \sum_{r=1}^{N_R}P_r  }{\sigma^2}$ be the per-hop SNRs for the first and second hops, respectively. Assuming that the CIRs are normalized such that $\sum_{l=1}^{L_{ir}}{\sigma_{ir,l}^2}=1$, $i\in\{A,B\}$, $r\in\{1,2,\ldots,N_R\}$, we have $\left|q_{ir,k}\right|\sim Rayleigh (\frac{1}{\sqrt{2}})$ and $\left|q_{ri,k}\right|\sim Rayleigh (\frac{1}{\sqrt{2}})$. By dropping the second term of the numerator of \eqref{eqn:SNR_B_2} and using $\gamma_{2}$ as the SNR for the second hop, the performance of the JBD scheme can be approximated by the performance of the single relay systems in \cite{Hasna_2003,Song_2010}.

Assuming BPSK modulation, the average probability of bit error at user B in the high SNR region can be approximated in terms of the per-relay SNR (i.e., $\gamma_1$) and the SNR of the second hop linking the relays to user B (i.e., $\gamma_2$) as
\begin{equation}\label{Pb_JBD}
P_{e,B} \approx \frac{1}{\gamma_1}+\frac{1}{2 \gamma_2}.
\end{equation}

We finally note that dropping the cross terms in the numerator of \eqref{eqn:SNR_B_2} has the advantage of mathematical tractability, and as the numerical examples will show later on, the approximation closely match the actual system performance, especially for high SNR values.

\section{The DSTC-Based Joint Blind-Differential (JBD-DSTC) Scheme} \label{sec:JBD-DSTC}
In multi-antenna single-way relay systems, distributed space-time coding (DSTC) was proposed in \cite{Hassibi_DSTC_2006} based on linear dispersion space-time codes (STCs) to mimic having an STC structure at the destination similar to the one obtained in multi-input single-output (MISO) systems that uses STCs. The system in \cite{Hassibi_DSTC_2006} assumes that there is CSI knowledge only at the destination. When there is no CSI knowledge, the differential DSTC can be used \cite{Jafarkhani_DDSTC_2008}.

In this section, we describe the proposed JBD-DSTC scheme based on differential DSTC transmission for a multi-relay TWR system in order to fully harness the inherent diversity advantage of this system. We consider a frame composed of $M$ blocks in which $T$ blocks are grouped together. There are $M_G$ groups in a frame where $M_G=M/T$, and the symbols over one subcarrier from the blocks of each group correspond to one space-time (ST) codeword.

\begin{figure}
\centering
 \includegraphics[clip, scale=1.7] {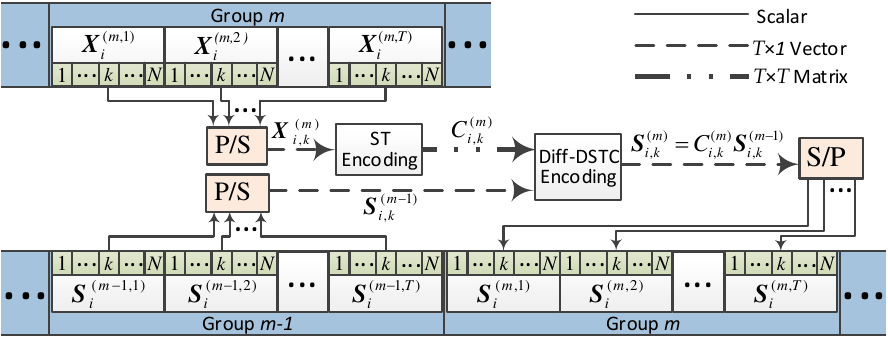}
\vspace{-0mm} \caption{Encoding process of the JBD-DSTC scheme at the $i$th user for the $T$ symbols over the $k$th subcarrier during the $m$th group. The green boxes represent the symbols on the $N$ subcarriers for the corresponding block and the  notations P/S and S/P denote parallel to serial and serial to parallel, respectively.}
\label{fig:JBD_DSTC_data}
\end{figure}

Fig. \ref{fig:JBD_DSTC_data} illustrates the encoding process at the $i$th user for the $T$ symbols over the $k$th subcarrier during the $m$th group. Note that $N$ parallel encoders are required for the entire $N$ subcarriers.
As shown in Fig. \ref{fig:JBD_DSTC_data}, the frequency-domain data-bearing vector of the $i$th user, $i\in\{A,B\}$, during the $t$th block of the $m$th group is denoted by ${  \bm X}_{i}^{(m,t)}$ where ${  \bm X}_{i}^{(m,t)}=\left[ {X}_{i,1}^{(m,t)} ,{X}_{i,2}^{(m,t)}, \ldots,{X}_{i,N}^{(m,t)}\right]^T$ and ${X}_{i,k}^{(m,t)} \in{\mathcal{A}_i}$. Prior to differential encoding, the vector of data symbols over the same subcarrier, $k$, and over all blocks of the same group, $m$, i.e., ${\bm X}_{i,k}^{(m)} = \left[ {X}_{i,k}^{(m,1)} ,{X}_{i,k}^{(m,2)}, \ldots,{X}_{i,k}^{(m,T)}\right]^T$, is encoded as a $T \times T$ unitary matrix $ {C}_{i,k}^{(m)} $. The structure of this matrix is designed such that it commutes with the linear dispersion matrices at the relays \cite{Jafarkhani_DDSTC_2008}. Let ${\cal C}$ denote the set of all possibilities of such matrices. Note that having a unitary structure preserves the transmit power at each user. 

Using differential DSTC (Diff-DSTC), each user differentially encodes the $T$ symbols on the $k^{th}$ subcarrier of the $T$ blocks belonging to the $m$th group as ${\bm S}_{i,k}^{(m)} = {C}_{i,k}^{(m)}  {\bm S}_{i,k}^{(m-1)}$, $m\in\{2,3,\ldots,M_G\}$ where ${\bm S}_{i,k}^{(m)}=\left[ {S}_{i,k}^{(m,1)} ,{S}_{i,k}^{(m,2)}, \ldots,{S}_{i,k}^{(m,T)}\right]^T$ and ${\bm S}_{i,k}^{(1)}$ is an arbitrary $T\times 1$ reference vector with elements from $\mathcal{A}_i$. 
Let ${  \bm S}_{i}^{(m,t)}=\left[ {S}_{i,1}^{(m,t)} ,{S}_{i,2}^{(m,t)}, \ldots,{S}_{i,N}^{(m,t)}\right]^T$. After performing IDFT, we obtain ${\bm s}_{i}^{(m,t)}=\left[ {s}_{i,1}^{(m,t)} ,{s}_{i,2}^{(m,t)}, \ldots,{s}_{i,N}^{(m,t)}\right]^T={\rm IDFT} ( {  \bm S}_{i}^{(m,t)})$.
The transmitted signal from the $i$th user during the $t$th block of the $m$th group, $i\in\{A,B\}$, is given by ${  \bm s}_{Tx,i}^{(m,t)}={  \bm s}_{Tx,i}^{(m,t)}=\left[ {s}_{Tx,i,1}^{(m,t)} ,{s}_{Tx,i,2}^{(m,t)}, \ldots,{s}_{Tx,i,N+N_{CP,1}}^{(m,t)}\right]^T = \sqrt{P_{i}} \zeta_1\left(  {\bm s}_{i}^{(m,t)} \right)$.
\subsection{Relay Processing}\label{subsec:JBD-DSTC_RelayProcessing}
After CP removal during the MAC phase at the $r$th relay, the received superimposed signal for the $t$th OFDM block of the $m$th group is given by
\begin{equation} 
\notag\begin{array}{lll}
 {\bm y}_{r}^{(m,t)}
 &=\sqrt{P_A}  H_{tl,Ar} \Psi_{d_{Ar}} {  \bm s}_{A}^{(m,t)} + \sqrt{P_B}  H_{tl,Br} \Psi_{d_{Br}} {  \bm s}_{B}^{(m,t)}
+{  \bm n}_{r}^{(m,t)},
\end{array}
\end{equation}
where ${\bm y}_{r}^{(m,t)}=\left[ {y}_{r,1}^{(m,t)} ,{y}_{r,2}^{(m,t)}, \ldots,{y}_{r,N}^{(m,t)}\right]^T$ and ${\bm n}_{r}^{(m,t)}$ is a CSCG random vector with mean ${\bm 0}_N$ and covarince matrix $\sigma_{ r}^2{I}_N$. To obtain the desired STC structure at the end-users, the $r$th relay processes $\left\{{y}_{r,n}^{(m,t)}\right\}_{t\in \{1,2,\ldots,T\}}$ to obtain ${  \bm s}_{r,n}^{(m)}$ as
\begin{eqnarray}
\notag \left[\begin{array}{c} s_{r,n}^{(m,1)}\\ s_{r,n}^{(m,2)}\\ \vdots \\s_{r,n}^{(m,T)} \end{array}\right] & = & A_r\left[\begin{array}{c}y_{r,n}^{(m,1)}\\ y_{r,n}^{(m,2)}\\ \vdots \\y_{r,n}^{(m,T)} \end{array}\right] +  B_r\left[\begin{array}{c}\eta\left(y_{r,n}^{(m,1)*}\right)\\ \eta\left(y_{r,n}^{(m,2)*}\right)\\ \vdots \\\eta\left(y_{r,n}^{(m,T)*}\right) \end{array}\right]
,
\end{eqnarray} %
$r=\{1,\ldots,N_R\}$, $n=\{1,\ldots,N\}$. The $T\times T$ relay dispersion matrices $A_r$ and $B_r$ are designed such that they commute with the data matrices, i.e., with ${C}_{i,k}^{(m)}$, while ensuring that the received signal at each user possesses the desired space-time block code (STBC) structure.

One simple design is introduced in \cite{Jafarkhani_DDSTC_2008} in which the relays are classified into two groups, ${\cal G}_1$ and ${\cal G}_2$. The $r$th relay falling into ${\cal G}_1$ uses a unitary matrix for $A_r$ and sets $B_r=0_{T\times T}$ while that falling into ${\cal G}_2$ sets $A_r=0_{T\times T}$ and uses a unitary matrix for $B_r$. According to this design, the relays' commutative property can be written as $ C  O_r=  O_r \widetilde{C}_{r} \hspace{2pt} \forall r $ where
\begin{equation}
\notag \begin{array}{rll}
 O_r =& \left\{   \begin{array}{ll}
A_r, &  r\in {\cal G}_1,\\
B_r, &  r\in {\cal G}_2,	\end{array}  \right.
 \end{array}
\hspace{-2pt} \text{and} \hspace{2pt}
\begin{array}{rll}
 \widetilde{C}_{r} =& \left\{   \begin{array}{ll}
C, &  r\in {\cal G}_1,\\
\mbox{$C$}^*, &  r\in {\cal G}_2.	\end{array}  \right.
 \end{array}
\end{equation}
Hence, we can write the set of all possible STC data matrices as
\begin{equation}
\notag {\cal C} = \left\{ C\left| C^H C= C C^H  =I_{T\times T}, C  O_r=  O_r \widetilde{C}_{r} \hspace{2pt} \forall r \right. \right\}.
\end{equation}
To simplify the estimation of the self-interference term, we impose another design criterion on the relay dispersion matrices, that is, all the matrices of the form $O_i^H O_j$, $i,j\in\{1,2,\ldots,N_R\}$, $i \neq j$, are hollow matrices, i.e., their diagonal entries are all zeros.

The $t$th transmitted block of the $r$th relay during the $m$th group is given by ${  \bm s}_{Tx,r}^{(m,t)} = \sqrt{P_r G_{r}} \zeta_2\left(  {\bm s}_{r}^{(m,t)} \right)$ where

${  \bm s}_{Tx,r}^{(m)}=\left[ {s}_{Tx,r,1}^{(m)} ,{s}_{Tx,r,2}^{(m)}, \ldots,{s}_{Tx,r,N+N_{CP,2}}^{(m)}\right]^T$ and
${\bm s}_{r}^{(m,t)}=\left[ {s}_{r,1}^{(m,t)} ,{s}_{r,2}^{(m,t)}, \ldots,{s}_{r,N}^{(m,t)}\right]^T$.

\subsection{Detection at the End-User}\label{subsec:JBD-DSTC_Detection}
By the end of the BC phase, and after removing the CP of length $N_{CP,2}$ at user B, the resulting consecutive $N$-sample OFDM blocks of the $t$th block, $t\in\{1,2,\ldots,T\}$, in the $m$th group, $m \in \left\lbrace 1,M_G \right\rbrace $, is denoted by ${\bm y}_{B}^{(m,t)}$. After performing DFT, the frequency-domain signal corresponding to ${\bm y}_{B}^{(m,t)}$ is ${\bm Y}_{B}^{(m,t)} = \left[ Y_{B,1}^{(m,t)},Y_{B,2}^{(m,t)},\ldots,Y_{B,N}^{(m,t)} \right]^T$ where ${\bm Y}_{B}^{(m,t)} = {\rm DFT} ({\bm y}_{B}^{(m,t)})$.
Let ${\bm V}_{B}^{(m,t)} = \left[ V_{B,1}^{(m,t)},V_{B,2}^{(m,t)},\ldots,V_{B,N}^{(m,t)} \right]^T $ denote the frequency-domain noise vector observed at user $B$ during the $t$th block of the $m$th group and let ${\bm Y}_{B,k}^{(m)}=\left[ Y_{B,k}^{(m,1)},Y_{B,k}^{(m,2)},\ldots,Y_{B,k}^{(m,T)} \right]^T$ denote the vector of received signals from all blocks of the $m$th group on the $k^{th}$ subcarrier. Similarly, define ${\bm V}_{B,k}^{(m)}=\left[ V_{B,k}^{(m,1)},V_{B,k}^{(m,2)},\ldots,V_{B,k}^{(m,T)} \right]^T$ and ${D}_{i,k}^{(m)} = \left[ O_1 \widetilde{\bm S}_{i,k,1}^{(m)},O_2 \widetilde{\bm S}_{i,k,2}^{(m)} ,\ldots, O_{N_R} \widetilde{\bm S}_{i,k,N_R}^{(m)} \right] $, $i\in\{A,B\}$ where  \begin{equation}
\notag \begin{array}{rlllll}
 \widetilde{\bm S}_{i,k,r}^{(m)}  & = & \left[  \widetilde{S}_{i,k,r}^{(m,1)} , \widetilde{S}_{i,k,r}^{(m,2)}, \ldots, \widetilde{S}_{i,k,r}^{(m,T)}\right]^T & = & \left\{   \begin{array}{ll}
{\bm S}_{i,k}^{(m)}, &  r\in {\cal G}_1,\\
\mbox{${\bm S}_{i,k}^{(m)}$}^*, &  r\in {\cal G}_2.	\end{array}  \right.
 \end{array}
\end{equation}
 $ \widetilde{\bm S}_{i,k,r}^{(m)}=\left[  \widetilde{S}_{i,k,r}^{(m,1)} , \widetilde{S}_{i,k,r}^{(m,2)}, \ldots, \widetilde{S}_{i,k,r}^{(m,T)}\right]^T$

Let $q_{ij,k}=\left[  H_{df,ij}\right]_{k,k}$. We can write ${\bm Y}_{B,k}^{(m)}$ as\footnote{An illustrative example for a dual-relay system is given in Appendix \ref{sec:Diff_appendixB}.} ${\bm Y}_{B,k}^{(m)} =  {D}_{B,k}^{(m)} {\bm \mu}_{B,k} +
 {D}_{A,k}^{(m)} {\bm \mu}_{A,k}   + {\bm V}_{B,k}^{(m)}$
where ${\bm \mu}_{i,k}$, $i\in\{A,B\}$, are $N_R \times 1$ channel-dependent vectors defined as
\begin{equation} \label{eqn:muikm}
{\bm \mu}_{i,k}=\left[\begin{array}{cc} \sqrt{P_{i1}} q_{1B,k} \widetilde{q}_{i1,k}  e^{-j\frac{ 2\pi \left( k-1 \right) \left( d_{1B}+\widetilde{d}_{i1} \right)}{N}} \\ \sqrt{P_{i2}} q_{2B,k} \widetilde{q}_{i2,k}  e^{-j\frac{ 2\pi \left( k-1 \right) \left( d_{2B}+\widetilde{d}_{i2} \right)}{N}} \\
 \vdots\\
 \sqrt{P_{iN_R}} q_{N_RB,k} \widetilde{q}_{iN_R,k}  e^{-j\frac{ 2\pi \left( k-1 \right) \left( d_{N_RB}+\widetilde{d}_{iN_R} \right)}{N}}
 \end{array}\right], 
\end{equation}
where
\begin{equation}
\notag \begin{array}{rll}
 \widetilde{q}_{ij,k} =& \left\{   \begin{array}{ll}
q_{ij,k}, & j\in {\cal G}_1, \\
q^*_{ij,k}, &   j\in {\cal G}_2,	\end{array}  \right.
\end{array}
\hspace{-2pt} \text{and} \hspace{2pt}
\begin{array}{rll}
 \widetilde{d}_{ij} =& \left\{   \begin{array}{ll}
d_{ij}, &   j\in {\cal G}_1, \\
-d_{ij}, &    j\in {\cal G}_2.
\end{array}  \right.
\end{array}
\end{equation}

For a sufficiently large $M$, we can obtain an estimate of ${\bm \mu}_{B,k}$, denoted by $\widehat{{\bm \mu}}_{B,k}$, as\footnote{The derivation of this result is outlined in Appendix \ref{sec:Diff_appendixC}.}
\begin{equation} \widehat{{\bm \mu}}_{B,k}
  \approx  \sum_{m=1}^{M}  \mbox{$ {D}_{B,k}^{(m)} $}^H {\bm Y}_{B,k}^{(m)} / (M T),
  \end{equation}
Note that unlike the JBD scheme, the JBD-DSTC scheme does not require the channel reciprocity assumption. Having obtained an estimate for ${\bm \mu}_{B,k}$, user $B$ can remove its estimated self-interference term, ${D}_{B,k}^{(m)} \widehat{{\bm \mu}}_{B,k} $ to obtain ${\bm Y}_{AB,k}^{(m)} \approx  {D}_{A,k}^{(m)} {\bm \mu}_{A,k}    +{\bm V}_{B,k}^{(m)}$.
Using the commutative property and the fact that ${\bm S}_{i,k}^{(m)}$ is differentially encoded, we can simplify ${\bm Y}_{AB,k}^{(m)}$ as


\begin{equation} \label{eqn:yBkmv_1_small_noSB}
\begin{array}{lll}
{\bm Y}_{AB,k}^{(m)} & \approx  & \left[ O_1 \widetilde{C}_{A,k,1}^{(m)} \widetilde{\bm S}_{A,k,1}^{(m)},O_2 \widetilde{C}_{A,k,2}^{(m)} \widetilde{\bm S}_{A,k,2}^{(m)},\ldots,O_{N_R} \widetilde{C}_{A,k,N_R}^{(m)} \widetilde{\bm S}_{A,k,N_R}^{(m)} \right]    \widehat{{\bm \nu}}_{k} +{\bm V}_{B,k}^{(m)} \\
&  \approx  & \left[C_{A,k}^{(m)}  O_1 \widetilde{\bm S}_{A,k,1}^{(m-1)},C_{A,k}^{(m)} O_2  \widetilde{\bm S}_{A,k,2}^{(m-1)},\ldots,C_{A,k}^{(m)} O_{N_R} \widetilde{\bm S}_{A,k,N_R}^{(m-1)} \right]   \widehat{{\bm \nu}}_{k} +{\bm V}_{B,k}^{(m)} \\
&  \approx  & C_{A,k}^{(m)} {\bm Y}_{AB,k}^{(m-1)} + \left( {\bm V}_{B,k}^{(m)} - C_{A,k}^{(m)} {\bm V}_{B,k}^{(m-1)} \right)   , \quad  m=2,3,\ldots,M_G
 \end{array}
\end{equation}
where
 \begin{equation}
\notag \begin{array}{rll}
 \widetilde{C}_{A,k,r}^{(m)} = & \left\{   \begin{array}{ll}
C_{A,k}^{(m)}, &  r\in {\cal G}_1,\\
\mbox{$C_{A,k}^{(m)}$}^*, &  r\in {\cal G}_2,	\end{array}  \right.
\end{array}
\end{equation}

Therefore, ${C}_{A,k}^{(m)}$ can be recovered at user $B$ using the following detection rule
	\begin{align}
 \widehat{C}_{A,k}^{(m)} &=\arg\,\underset{C\in\mathcal{C}   }{\min} \left\Vert {\bm Y}_{AB,k}^{(m)} -  C  {\bm Y}_{AB,k}^{(m-1)} \right\Vert ^2  , \quad  m=2,3,\ldots,M_G.\label{Eq:Chat}
  \end{align}
Note that if $C$ has an STBC structure, then the above equation can be easily decoupled, which allows fast symbol-wise ML detection. Similar to the JBD scheme, employing ideas based on multiple-symbol differential detection, which in this case involves the joint detection of the $M_G$ data matrices, promises significant performance improvements, which comes at the expense of increased receiver complexity. 


\subsection{Performance Analysis} \label{PEP_JBD_DSTC}
Inspired by the results obtained in \cite{Jafarkhani_DDSTC_2008} for single-way differential DSTC, we can write the pairwise error probability of mistaking ${C}_{A,k}^{(m)}$ by ${C'}_{A,k}^{(m)}$, i.e., $\mathbb{P}\left( {C}_{A,k}^{(m)}  \to  {C'}_{A,k}^{(m)}   \right) $ in the two-way relaying scheme under consideration. Let $\sigma_i^2=\sigma_r^2=\sigma^2$ $\forall$ $i\in\{A,B\}$, $r\in\{1,2,\ldots,N_R\}$. Assuming that the CIRs are normalized such that $\sum_{l=1}^{L_{ir}}{\sigma_{ir,l}^2}=1$, $i\in\{A,B\}$, $r\in\{1,2,\ldots,N_R\}$, the PEP, averaged over channel realizations, can be approximately upper bounded for large SNR values as
\begin{equation} \label{eqn:PEP_JBD_DSTC}
\mathbb{P}\left( {C}_{A,k}^{(m)}  \to  {C'}_{A,k}^{(m)}   \right)   \vspace{-0pt} 
       \begin{array}{c}
         \\ [-35pt] \hspace{-1pt} \displaystyle{\rotsim}  \\ [-17pt] <
         \end{array}
         \frac{\left(  \frac{16 N_R\log \Omega}{ \Omega T}\right)^{N_R}} {\Delta\left({C}_{A,k}^{(m)} , {C'}_{A,k}^{(m)}\right)}
\end{equation} 
where $\Omega =  \sqrt{\frac{2}{T}}\frac{\left( P_A+ P_B + \sigma^2 \right)\sum_{r=1}^{N_R}P_{r}}{\sigma^2}$ and $\Delta( C, C') = \det\left( (C-C')^* (C-C') \right)$ gives an indication of the distance between $C$ and $C'$.

With the assumption that $\frac{\sum_{r=1}^{N_R}P_{r}}{\sigma^2} \gg 1$, the JBD-DSTC scheme can achieve a diversity of $N_R \left( 1 - \frac{\log \log \Omega}{ \log \Omega }\right) $.

\section{Numerical Results} \label{sec:performanceecval_Diffsmall} 
As an example, we consider a frequency-selective Rayleigh fading channel with three taps defined by $\left\lbrace  \sigma{_{ir,l}} \right\rbrace_{l\in\{1,2,3\} } = \frac{[1,\hspace{3pt}0.8,\hspace{3pt}0.6 ]}{\sqrt{2}}$, $i\in\{A,B\}$, $r\in\{1,2,\ldots,N_R\}$, $N=64$ subcarriers and total bandwidth of $8$ kHz. The selection of the available bandwidth is consistent with, for example, underwater acoustic communications. The SNR at user $i$ while detecting the signal of user $i'$ is defined as $ SNR_i=\left( G_1 +G_2 \right) P_{i'} /\sigma_{i,eff}^2$,  $i,i'\in\{A,B\}$, $i'\neq i$ where $\sigma_{i,eff}^2=G_1 \sigma_{1}^2+G_2 \sigma_{2}^2 +\sigma_{i}^2$ is the effective noise variance at user $i$. Unless stated otherwise, Quadrature PSK (QPSK) is used and $\sigma_{B}^2=\sigma_{1}^2=\sigma_{2}^2=\sigma^2$. We further assume that $N_R=2$, $P_A=1$, $G_1=G_2=1$, $d_{A1}=5$, $d_{B1}=14$, $d_{A2}=3$, $d_{B2}=9$, $d_{1B}=14$ and $d_{2B}=9$. For the JBD-DSTC scheme, two blocks per group ($T=2$) is assumed, and we adopt the dispersion matrices designed in \cite{Jafarkhani_DDSTC_2008}.

\begin{figure}
\centering\includegraphics[ clip, width=\matfigwidth]{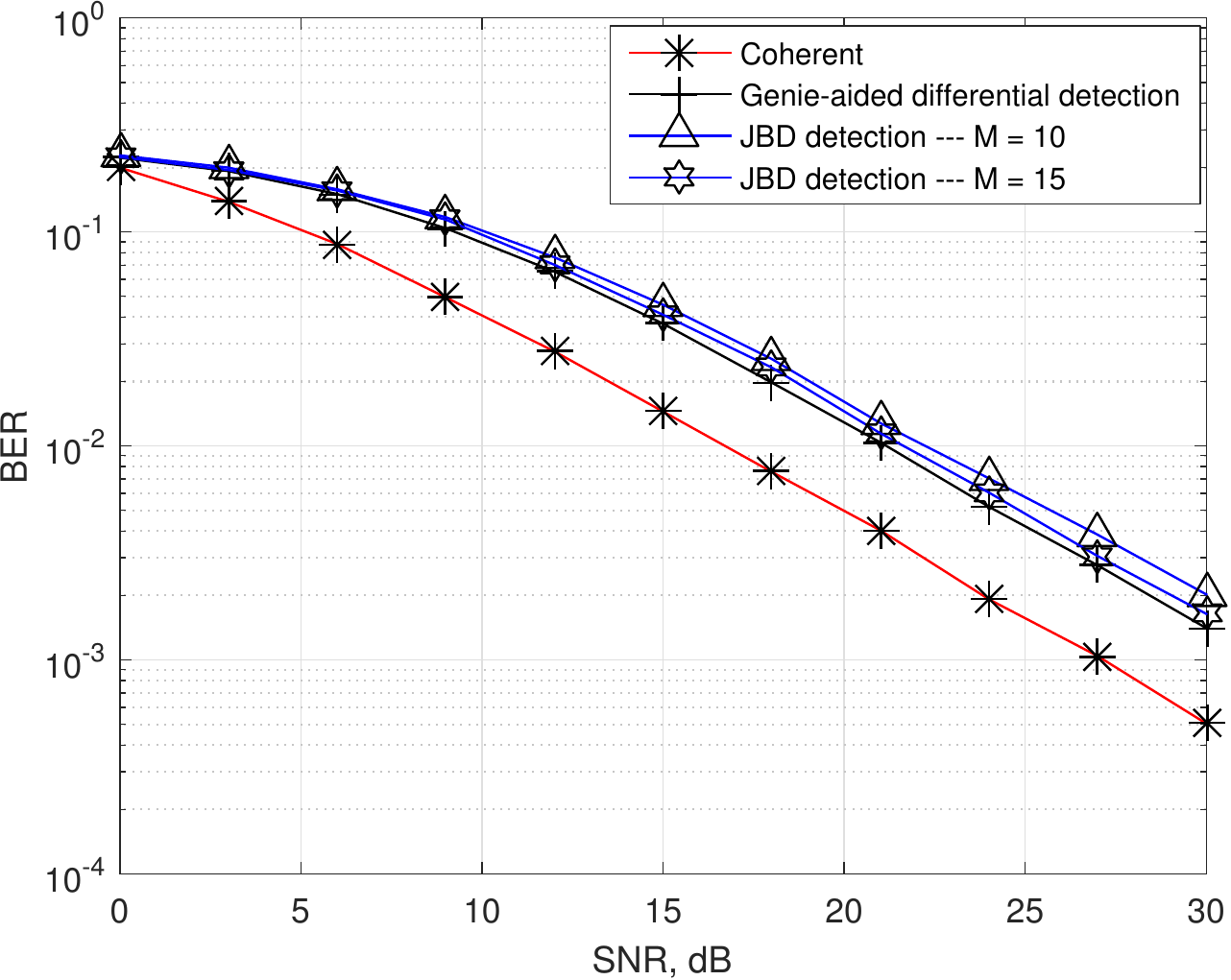}
\vspace{-0mm}  \caption{BER performance of the JBD detector and the coherent detector.} \label{fig:JBD_M_all}
\vspace{-0mm}
\end{figure}

In Fig. \ref{fig:JBD_M_all}, we compare the BER performance of the JBD detector with that of the coherent detector. Clearly, the coherent scheme outperforms the differential scheme by almost $3$ dB which is an expected result. We also plot the performance of a genie-aided differential detector that assumes the knowledge of $\mu_{1,k}$ and $\mu_{2,k}$ $\forall k$, at user $B$ and the knowledge of $\nu_{1,k}$ and $\nu_{2,k}$ $\forall k$, at user $A$, and hence self-interference is perfectly removed. As seen in Fig. \ref{fig:JBD_M_all}, if $15$ blocks are assumed, the performances of the two schemes match closely, which shows the accuracy of the parameters estimation. Furthermore,it shows that our proposed scheme still performs close to the genie-aided case even if the number of blocks is reduced from to $10$. Similar results are observed for JBD-DSTC. 

\begin{figure}
\centering\includegraphics[ clip, width=\matfigwidth]{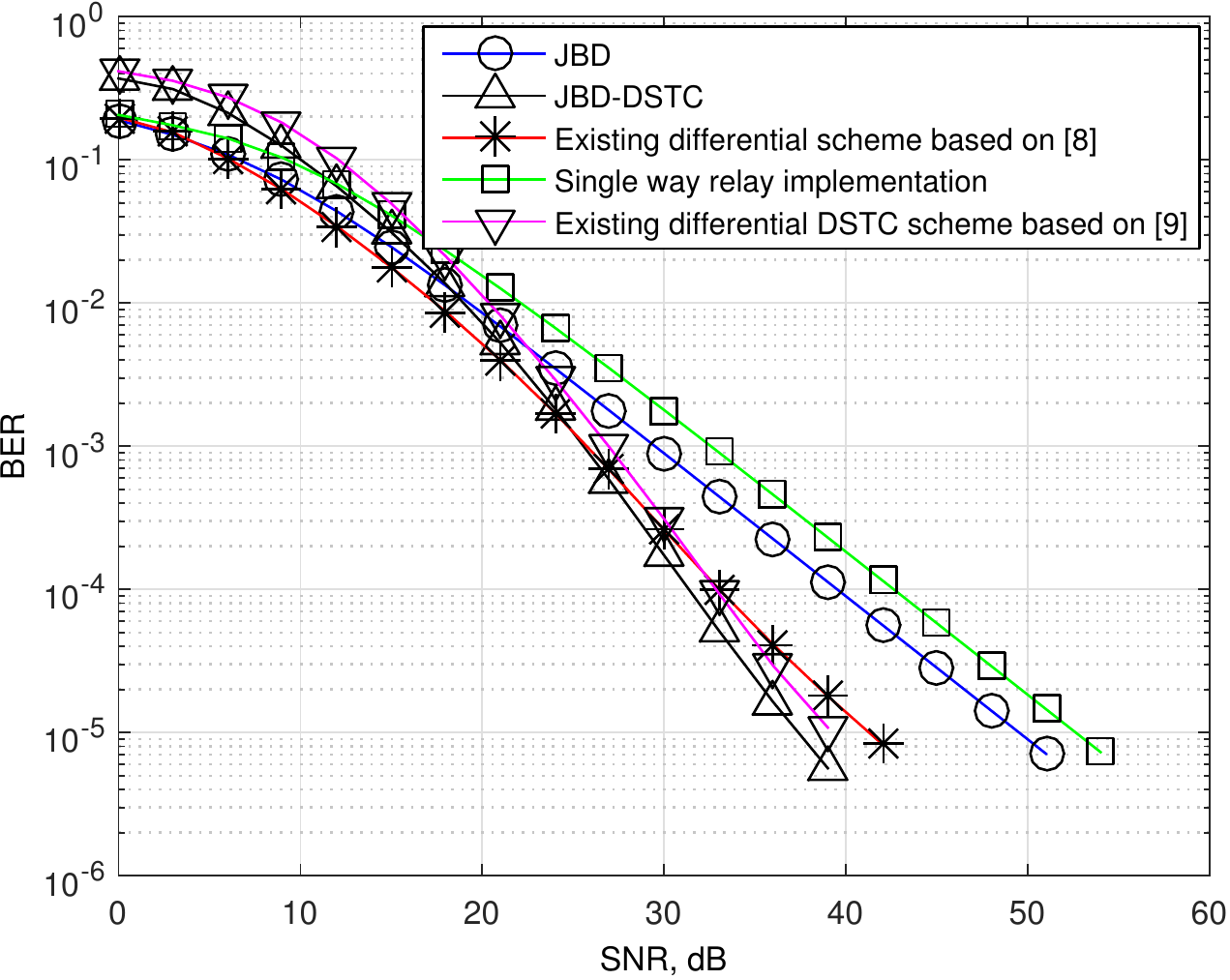}
\vspace{-0mm}   \caption{BER performance of the proposed schemes and some existing schemes ($M=200$).} \label{fig:BER_all_M200_N64_Lh3_allfdts_minerror200}
\vspace{-0mm}
\end{figure}

In Fig. \ref{fig:BER_all_M200_N64_Lh3_allfdts_minerror200}, we compare our proposed schemes to two existing differential-based TWR schemes along with the conventional single-way relay (SWR) implementation when the channel is quasi-static. For SWR implementation, four phases of transmission are required and hence we use QPSK rather than BPSK as in the TWR schemes to unify the transmission rate. For the two schemes in \cite{utkovski_2009_distributed,Debbah2010}, we properly extend their proposals to the multicarrier case to perform the comparison. Clearly, the JBD scheme outperforms the JBD-DSTC scheme for low SNR values (below $17$ dB for this example) while the opposite happens for higher SNR values as JBD-DSTC achieves the full diversity order of 2. In fact, the JBD-DSTC scheme outperforms all the other considered schemes in the high SNR region (greater than $25$ dB here). Specifically, it outperforms the scheme in \cite{utkovski_2009_distributed}, the one in \cite{Debbah2010}, the JBD scheme and the SWR system by about $1.5$ dB, $1.7$ dB, $8.2$ dB and $11.3$ dB, respectively, at a BER of about $10^{-4}$. Specifically, we attribute the improvement over the scheme in \cite{utkovski_2009_distributed}, which is also based on differential DSTC, to the fact that the detector in \cite{utkovski_2009_distributed} uses estimates of the partner's previous symbol (in addition the currently received signal) to detect the partner's current symbol which causes error propagation. In our scheme, on the other hand, the detection of the current symbol is independent from the previous symbol.

We can note from Fig. \ref{fig:BER_all_M200_N64_Lh3_allfdts_minerror200} that the scheme in \cite{Debbah2010} which is based on relay selection diversity performs better than all other proposals for small SNR values the (below $25$ dB for this example). However, it imposes a transmission overhead as it requires sending a sufficient number of pilot symbols to aid in assigning specific subcarrier(s) to the relay that minimizes the total symbol error rate of the users over this (those) subcarrier(s), and after that, additional feedback is required to broadcast the indices of the subcarriers that each relay should handle. Furthermore, unlike our schemes, the relays are required to perform DFT and IDFT to enable filtering out all subcarriers except the ones assigned to each one of them. 

\begin{figure}
\centering\includegraphics[ clip, width=\matfigwidth]{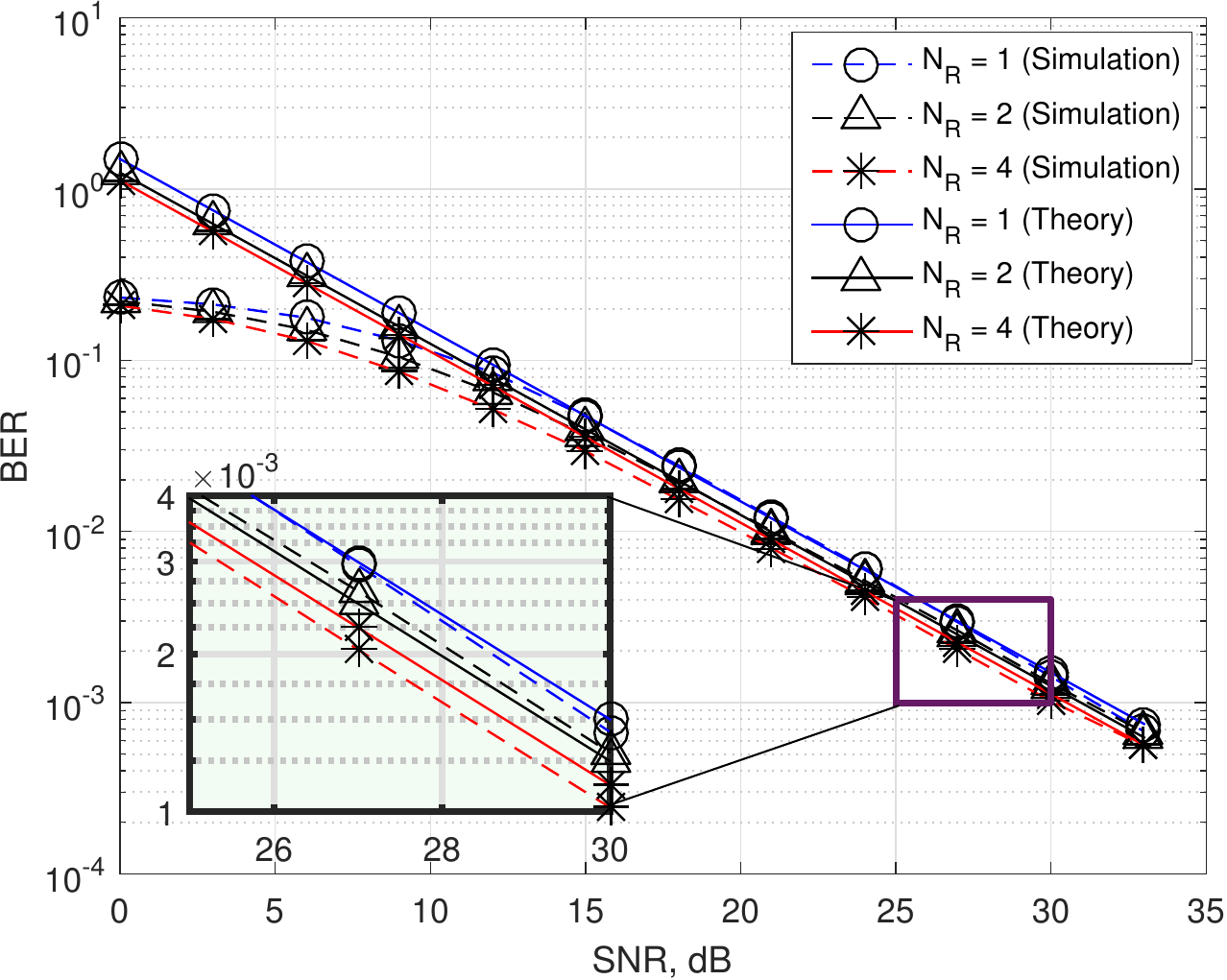}
\vspace{-0mm}  \caption{Comparison between analytical and simulation performance results for the JBD detector ($M=200$).} \label{fig:JBD_NR}
\vspace{-0mm}
\end{figure}

Fig. \ref{fig:JBD_NR} compares the analytical and the simulation performance results for the JBD scheme using BPSK modulation. Herein, the power at the relay is normalized as explained in Section \ref{Pe_JBD} and the transmit power of the $r$th relay, $P_r$, $r\in\{1,2,\ldots,N_R\}$ is set to unity. Fig. \ref{fig:JBD_NR} shows a close match between simulation results and the analytical $P_b$ (as in \eqref{Pb_JBD}) in the high SNR region (greater than $15$ dB for this example) for various number of relays.

\begin{figure}
\centering\includegraphics[ clip, width=\matfigwidth]{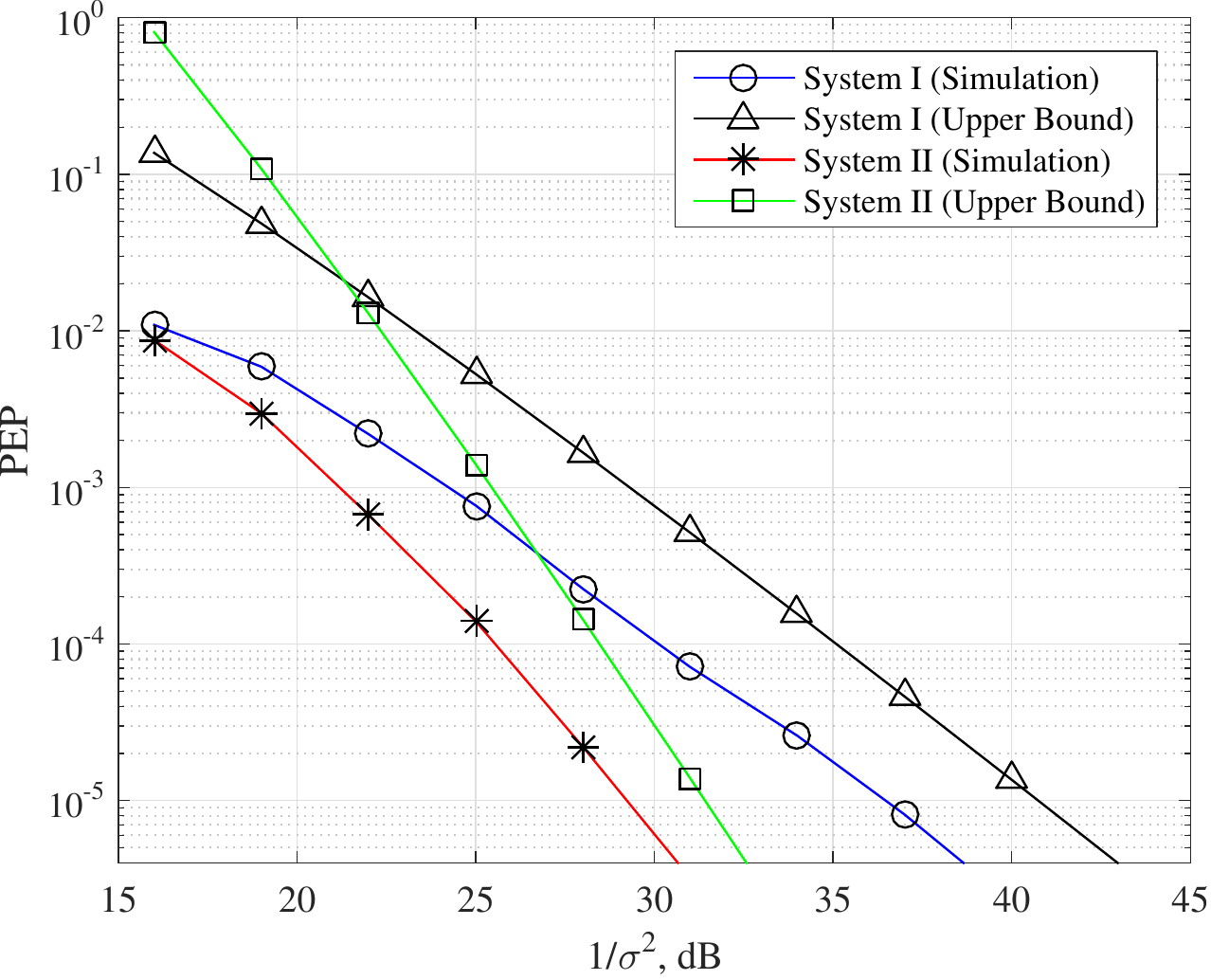}
\vspace{-0mm}  \caption{Comparison between analytical PEP upper bound and simulation results for the JBD-DSTC detector ($M=400$).} \label{fig:JBD_DSTC_NR}
\vspace{-0mm}
\end{figure}

In Fig. \ref{fig:JBD_DSTC_NR}, we compare between the analytical PEP upper bound of the JBD-DSTC detector in \eqref{eqn:PEP_JBD_DSTC} to the estimated PEP obtained from Monte Carlo simulations. We consider two scenarios for the number of relays, namely 2 and 4 which are implemented using groups of sizes $T=2$ and $T=4$, respectively. Here, we use BPSK modulation and hence we can adopt the square real orthogonal dispersion matrices proposed in \cite{Tarokh_1999}. The following summarizes the structure of the data matrices and the dispersion matrices for the two scenarios:
\begin{description}
  \item[\textbf{System I}] \hspace{20pt}
  \begin{equation} 
 {C}_{i,k}^{(m)}=\frac{1}{\sqrt{\left| {X}_{i,k}^{(m,1)} \right|^2+\left| {X}_{i,k}^{(m,2)} \right|^2}}\left[\begin{array}{cc} {X}_{i,k}^{(m,1)} & -{X}_{i,k}^{(m,2)}\\ {X}_{i,k}^{(m,2)} & {X}_{i,k}^{(m,1)} \end{array} \right],
\end{equation}
  \begin{equation} \label{eqn:relaymatrices}
A_1=I_2  \text{ and } A_2=\left[\begin{array}{cc}0 & -1\\ 1 & 0\end{array}\right].
\end{equation}
 \item[\textbf{System II}] \hspace{50pt}
 \begin{equation} 
 {C}_{i,k}^{(m)}=\frac{1}{\sqrt{\sum_{j=1}^{4}\left| {X}_{i,k}^{(m,j)} \right|^2}}\left[\begin{array}{cccc} {X}_{i,k}^{(m,1)} & -{X}_{i,k}^{(m,2)} & -{X}_{i,k}^{(m,3)} & -{X}_{i,k}^{(m,4)}\\
 {X}_{i,k}^{(m,2)} & {X}_{i,k}^{(m,1)} & {X}_{i,k}^{(m,4)} & -{X}_{i,k}^{(m,3)} \\
 {X}_{i,k}^{(m,3)} & -{X}_{i,k}^{(m,4)} & {X}_{i,k}^{(m,1)} & {X}_{i,k}^{(m,2)} \\
 {X}_{i,k}^{(m,4)} & {X}_{i,k}^{(m,3)} & -{X}_{i,k}^{(m,2)} & {X}_{i,k}^{(m,1)} \end{array} \right],
\end{equation}
\begin{equation} \label{eqn:relaymatrices}
A_1=I_4 \hspace{-0pt}, \hspace{2pt} A_2=\left[\begin{array}{cccc}0 & -1 &0 &0\\ 1 & 0&0&0\\0&0&0&-1\\0&0&1&0\end{array}\right]
\hspace{-0pt}, \hspace{2pt} A_3=\left[\begin{array}{cccc}0 & 0 &-1 &0\\ 0 & 0&0&1\\1&0&0&0\\0&-1&0&0\end{array}\right]
\text{ and } A_4=\left[\begin{array}{cccc}0 & 0 &0 &-1\\ 0 & 0&-1&0\\0&1&0&0\\1&0&0&0\end{array}\right].
\end{equation}
\end{description}
Note that for the two systems, $B_r=0_{T\times T}$, $r\in\{1,2,\ldots,N_R\}$.

Let ${\bm X}_{i,k}^{(m)}=\left[ {X}_{i,k}^{(m,1)} ,{X}_{i,k}^{(m,2)}, \ldots,{X}_{i,k}^{(m,T)}\right]^T$ denote data samples corresponding to the data matrix ${C}_{i,k}^{(m)}$. Similarly, ${\bm X'}_{i,k}^{(m)}$ corresponds to ${C'}_{i,k}^{(m)}$. to maintain fairness between the two scenarios, we consider ${\bm X}_{i,k}^{(m)}=\left[ 1,1 \right]^T$ and ${\bm X'}_{i,k}^{(m)}=\left[ -1,-1\right]^T$ for System I, while for System II, ${\bm X}_{i,k}^{(m)}=\left[ 1,1,1,1 \right]^T$ and ${\bm X'}_{i,k}^{(m)}=\left[ -1,-1,1,1 \right]^T$. Note that for the two scenarios, $\Delta\left({C}_{A,k}^{(m)} , {C'}_{A,k}^{(m)}\right)=16$. For Fig. \ref{fig:JBD_DSTC_NR}, we assume $P_A=1$, $P_r=\frac{1}{N_R}$ and $G_r=\left( P_A+ P_B + \sigma_r^2 \right)^{-1}$, $r\in\{1,2,\ldots,N_R\}$. Fig. \ref{fig:JBD_DSTC_NR} shows the validity of the upper bound and it also shows that the diversity is about 2 and 4 for systems I and II, respectively, as the PEP drops about 2 and 4 orders of magnitude, respectively, for an SNR increase of 10 dB.

\vspace{-.55cm}
\section{Conclusions} \label{sec:conclusions_Diffsmall} 
This paper proposed two schemes for differential asynchronous MR-TWR systems in frequency-selective fading channels in which neither the knowledge of the CSI nor the propagation delays is required. An advantage of these schemes is that the relays are only required to perform simple operations on the received (overlapped) signals, e.g., complex conjugation and time-reversal. Also, after estimating the channel-dependent parameters, only a simple symbol-wise detection rule is required. Through numerical simulations, it is observed that the proposed schemes are superior to the existing ones in the literature. The paper has also provided analytical error probability results for the proposed schemes that matched the results of Monte Carlo simulations.

\newpage

\appendices
\section{Simplification of $Y_{B,k}^{(m)}$ for the JBD scheme} \label{sec:Diff_appendixA}
After DFT, the $m$th block of the effective signal in frequency-domain can be written as
\begin{equation} \label{eqn:YB}
\begin{array}{lll}
 {  \bm Y}_{B}^{(m)}
	&\overset{(a)}{=}& \displaystyle \sum_{i\in\{A,B\}} \sum_{r=1}^{N_R} \sqrt{P_{ir}} F H_{tl,rB} F^H F \Psi_{d_{rB}} F^H F \eta\left( H_{tl,ir}^{*} \Psi_{d_{ir}}^* {\bm s}_{i}^{(m)*}\right)
  +F {\bm v}_{B}^{(m)}\\
		 &\overset{(b)}{=}& \displaystyle \sum_{i\in\{A,B\}} \sum_{r=1}^{N_R}
  \sqrt{P_{ir}} F H_{tl,rB} F^H F \Psi_{d_{rB}} F^H \left(F H_{tl,ir} \Psi_{d_{ir}} {\bm s}_{i}^{(m)}\right)^*
  +{\bm V}_{B}^{(m)}\\
		 &=&  \displaystyle \sum_{i\in\{A,B\}} \sum_{r=1}^{N_R}\sqrt{P_{ir}} F H_{tl,rB} F^H F \Psi_{d_{rB}} F^H \left(F H_{tl,ir} F^H F \Psi_{d_{ir}} F^H F {\bm s}_{i}^{(m)}\right)^*
  + {\bm V}_{B}^{(m)}\\
   &=&  \displaystyle \sum_{i\in\{A,B\}} \sum_{r=1}^{N_R} \sqrt{P_{ir}} H_{df,rB}^{(m)} \Psi_{F,d_{rB}} \left( H_{df,ir}^{(m)} \Psi_{F,d_{ir}}  {\bm S}_{i}^{(m)}\right)^*
  + {\bm V}_{B}^{(m)}
 \end{array}
\end{equation}
where ${  \bm V}_{B}^{(m)}=F {  \bm v}_{B}^{(m)}$, $\Psi_{F,d} = F\Psi_{d}F^H$ and (a) follows from the fact that the DFT matrix is a unitary matrix, i.e. $F^HF=FF^H=I_N$ where $I_N$ is the size-$N$ identity matrix. The equality (b) follows from the fact that conjugation along with reversal in time-domain results in conjugation in frequency-domain, i.e. $F \eta\left( {\bm x^*}\right)=\left( F {\bm x}\right)^*$. 

In case of block fading or of course quasi-static, which is our assumption here, $H_{tl,ir}$ have a circulant structure causing $H_{df,ir}$ to be diagonal which means no inter-carrier interference (ICI) is present. When the channel is time-varying within the same OFDM block, neither $H_{tl,ir}$ will be circulant nor will $H_{df,ir}$ be diagonal, which means that the subcarrier orthogonality is lost, giving rise to ICI.

It is clear to see that due to the different time delays experienced by the components of the signal in \eqref{eqn:yB}, different circular shifts resulted. Since having a delay of $n$ samples in the time domain causes the $k^{th}$ subcarrier to have a phase shift of $e^{-j2 \pi n(k-1)/N}$, $k\in\{1,2,\ldots,N\}$, we can write the received signal on the $k^{th}$ subcarrier as
\begin{eqnarray} \label{eqn:yBk_small_1}
\nonumber Y_{B,k}^{(m)} &= &  \displaystyle \sum_{i\in\{A,B\}} \sum_{r=1}^{N_R} \sqrt{P_{ir}} \left[H_{df,rB}^{(m)}\right]_{k,k}   \left[ H_{df,ir}^{(m)}\right]_{k,k}^* e^{-j\frac{ 2\pi \left( k-1 \right) \left( d_{rB}-d_{ir} \right)}{N}}  {{S}_{i,k}^{(m)}}^* +
{V}_{B,k}^{(m)},
\end{eqnarray}

Since we assumed the channels to be reciprocal, then for all $i\in\{A,B\}$, $r\in\{1,2\}$, $H_{df,ir}=H_{df,ri}$. We also assume that $d_{ri}=d_{ir}$, $r\in\{1,2\}$, $i\in\{A,B\}$. Therefore, the received signal on the $k^{th}$ subcarrier during the $m$th block can be written as $Y_{B,k}^{(m)} =  \mu_{k} { { S}_{B,k}^{(m)}}^* +
 \nu_{k}  { { S}_{A,k}^{(m)}}^* +{V}_{B,k}^{(m)}$.

\section{Illustrative example for the JBD-DSTC scheme: Dual-relay case}  \label{sec:Diff_appendixB}

To clearly illustrate the resulting DSTC structure, we consider the case of having two relays ($N_R=2$) and using two blocks per group ($T=2$). For this case, we adopt the dispersion matrices design in \cite{Jafarkhani_DDSTC_2008} that results in Alamouti's code structure. Specifically, the relays' matrices are chosen as
\begin{equation} \label{eqn:relaymatrices}
A_1=\left[\begin{array}{cc}1 & 0\\ 0 & 1\end{array}\right]\hspace{-2pt}, \hspace{2pt} B_1=0_{T\times T},\hspace{2pt} A_2=0_{T\times T} \text{ and } B_2=\left[\begin{array}{cc}0 & -1\\ 1 & 0\end{array}\right].
\end{equation} 
Interestingly, for the case of $N_R=2$ and $T=2$, it was found in \cite{Jafarkhani_DDSTC_2008} that a space-time codeword, $ C$, satisfies the commutative property if and only if it follows the $2\times 2$ Alamouti structure. Hence, $ {C}_{i,k}^{(m)}$ is constructed as
\begin{equation} 
 {C}_{i,k}^{(m)}=\frac{1}{\sqrt{\left| {X}_{i,k}^{(m,1)} \right|^2+\left| {X}_{i,k}^{(m,2)} \right|^2}}\left[\begin{array}{cc} {X}_{i,k}^{(m,1)} & -\mbox{${X}_{i,k}^{(m,2)}$}^*\\ {X}_{i,k}^{(m,2)} & \mbox{${X}_{i,k}^{(m,1)}$}^* \end{array} \right].
\end{equation}

After removing the CP of length $N_{CP,2}$ at user B, the resulting two consecutive $N$-sample OFDM blocks of the $m$th group, $m \in \left\lbrace 1,M_G \right\rbrace $, can be written as
\begin{equation} \label{eqn:yB_1}
\begin{array}{lll}
 {\bm y}_{B}^{(m,1)}
 =& \sqrt{P_{A1}}  H_{tl,1B} \Psi_{d_{1B}}  H_{tl,A1} \Psi_{d_{A1}} {\bm s}_{A}^{(m,1)}
 - \sqrt{P_{A2}}   H_{tl,2B} \Psi_{d_{2B}} \eta\left( H_{tl,A2}^{*} \Psi_{d_{A2}}^* {\bm s}_{A}^{(m,2)*}\right)\\
 & \,+ \sqrt{P_{B1}} H_{tl,1B}  \Psi_{d_{1B}} H_{tl,B1} \Psi_{d_{B1}} {\bm s}_{B}^{(m,1)}
     - \sqrt{P_{B2}}   H_{tl,2B} \Psi_{d_{2B}}\eta\left( H_{tl,B2}^{*} \Psi_{d_{B2}}^* {\bm s}_{B}^{(m,2)*}\right)  +{  \bm v}_{B}^{(m,1)},
 \end{array}
\end{equation}
\begin{equation} \label{eqn:yB_2}
\begin{array}{lll}
 {\bm y}_{B}^{(m,2)}
 =& \sqrt{P_{A1}}  H_{tl,1B} \Psi_{d_{1B}}  H_{tl,A1} \Psi_{d_{A1}} {\bm s}_{A}^{(m,2)}
 + \sqrt{P_{A2}}   H_{tl,2B} \Psi_{d_{2B}} \eta\left( H_{tl,A2}^{*} \Psi_{d_{A2}}^* {\bm s}_{A}^{(m,1)*}\right)\\
 & \,+ \sqrt{P_{B1}} H_{tl,1B}  \Psi_{d_{1B}} H_{tl,B1} \Psi_{d_{B1}} {\bm s}_{B}^{(m,2)}
     + \sqrt{P_{B2}}   H_{tl,2B} \Psi_{d_{2B}}\eta\left( H_{tl,B2}^{*} \Psi_{d_{B2}}^* {\bm s}_{B}^{(m,1)*}\right)  +{  \bm v}_{B}^{(m,2)},
 \end{array}
\end{equation} 
where ${\bm v}_{B}^{(m,t)}$ represents length-$N$ effective noise vector at user B during the $t$th block of the $m$th group whose entries are AWGN random variables with zero mean and variance of $\sigma_{ B}^2$.

After performing DFT, the frequency-domain signal corresponding to the first block of the $m$th group can be written as
   \begin{IEEEeqnarray}{rCl}
 {  \bm Y}_{B}^{(m,1)}
	 &=&
\sqrt{P_{A1}} F H_{tl,1B} F^H F \Psi_{d_{1B}} F^H F H_{tl,A1} F^H F \Psi_{d_{A1}} {\bm s}_{A}^{(m,1)} \notag\\ &&
 - \sqrt{P_{A2}}  F H_{tl,2B} F^H F \Psi_{d_{2B}} F^H F \eta\left( H_{tl,A2}^{*}  \Psi_{d_{A2}}^*  {\bm s}_{A}^{(m,2)*}\right)\notag \\ \,
 & &\,+ \sqrt{P_{B1}} F H_{tl,1B}  F^H F \Psi_{d_{1B}} F^H F H_{tl,B1} F^H F \Psi_{d_{B1}} F^H F {\bm s}_{B}^{(m,1)} \notag\\ &&
     - \sqrt{P_{B2}}  F H_{tl,2B} F^H F \Psi_{d_{2B}} F^H F \eta\left( H_{tl,B2}^{*} \Psi_{d_{B2}}^* {\bm s}_{B}^{(m,2)*}\right)  +{  \bm V}_{B}^{(m,1)}
	\notag \\
		 &=&
\sqrt{P_{A1}} H_{df,1B} \Psi_{F,d_{1B}}  H_{df,A1} \Psi_{F,d_{A1}}  {\bm S}_{A}^{(m,1)}  \notag\\ && \,
 - \sqrt{P_{A2}}  H_{df,2B} \Psi_{F,d_{2B}} \left( H_{df,A2} \Psi_{F,d_{A2}} {\bm S}_{A}^{(m,2)}\right)^*\notag \\
 & &\,+ \sqrt{P_{B1}} H_{df,1B} \Psi_{F,d_{1B}} H_{df,B1}\Psi_{F,d_{B1}} {\bm S}_{B}^{(m,1)}  \notag\\ && \,
     - \sqrt{P_{B2}} H_{df,2B} \Psi_{F,d_{2B}} \left( H_{df,B2} \Psi_{F,d_{B2}} {\bm S}_{B}^{(m,2)}\right)^*+{\bm V}_{B}^{(m,1)}
\label{eqn:YB_1}
 \end{IEEEeqnarray} 
where ${\bm V}_{B}^{(m,t)} =F{\bm v}_{B}^{(m,t)}$. Similarly, we can write ${\bm Y}_{B}^{(m,2)}$ for the second block as
 \begin{equation} \label{eqn:YB_2}
\begin{array}{lll}
 {\bm Y}_{B}^{(m,2)}
		 &=&
\sqrt{P_{A1}} H_{df,1B} \Psi_{F,d_{1B}}  H_{df,A1} \Psi_{F,d_{A1}}  {\bm S}_{A}^{(m,2)} \\
  & &
 + \sqrt{P_{A2}}  H_{df,2B} \Psi_{F,d_{2B}} \left( H_{df,A2} \Psi_{F,d_{A2}} {\bm S}_{A}^{(m,1)}\right)^*\\
 & &\,+ \sqrt{P_{B1}} H_{df,1B} \Psi_{F,d_{1B}} H_{df,B1}\Psi_{F,d_{B1}} {\bm S}_{B}^{(m,2)} \\
  & &
     + \sqrt{P_{B2}} H_{df,2B} \Psi_{F,d_{2B}} \left( H_{df,B2} \Psi_{F,d_{B2}} {\bm S}_{B}^{(m,1)}\right)^*+{  \bm V}_{B}^{(m,2)}
	\\
 \end{array}
\end{equation}

With ${\bm Y}_{B,k}^{(m)}=\left[ Y_{B,k}^{(m,1)},Y_{B,k}^{(m,2)} \right]^T$ and ${\bm V}_{B,k}^{(m)}=\left[ V_{B,k}^{(m,1)},V_{B,k}^{(m,2)} \right]^T$, we can write ${\bm Y}_{B,k}^{(m)}$ as
\begin{equation} \label{eqn:yBkvec_small}
{\bm Y}_{B,k}^{(m)} =  {D}_{B,k}^{(m)} {\bm \mu}_{B,k} +
 {D}_{A,k}^{(m)} {\bm \mu}_{A,k}   + {\bm V}_{B,k}^{(m)},
\end{equation}
where
\begin{equation} \label{eqn:Dik}
{D}_{i,k}^{(m)}=\left[\begin{array}{cc} {{S}_{i,k}^{(m,1)}} & -{{S}_{i,k}^{(m,2)}}^*\\ {{S}_{i,k}^{(m,2)}} & {{S}_{i,k}^{(m,1)}}^*\end{array}\right], \quad i\in\{A,B\},
\end{equation}


\begin{equation} \label{eqn:mukm}
{\bm \mu}_{B,k}^{(m)}=\left[\begin{array}{cc} \sqrt{P_{B1}} \left[  H_{df,1B}^{(m)}\right]_{k,k} \left[H_{df,B1}^{(m)} \right]_{k,k} e^{-j\frac{ 2\pi \left( k-1 \right) \left( d_{1B}+d_{B1} \right)}{N}} \\ \sqrt{P_{B2}}\left[  H_{df,2B}^{(m)}\right]_{k,k}  \left[H_{df,B2}^{(m)} \right]_{k,k}^* e^{-j\frac{ 2\pi \left( k-1 \right) \left( d_{2B}-d_{B2} \right)}{N}} \end{array}\right], 
\end{equation}
and
\begin{equation} \label{eqn:nukm}
{\bm \mu}_{A,k}^{(m)}=\left[\begin{array}{cc} \sqrt{P_{A1}} \left[  H_{df,1B}^{(m)}\right]_{k,k} \left[H_{df,A1}^{(m)} \right]_{k,k} e^{-j\frac{ 2\pi \left( k-1 \right) \left( d_{1B}+d_{A1} \right)}{N}} \\ \sqrt{P_{A2}}\left[  H_{df,2B}^{(m)}\right]_{k,k}  \left[H_{df,A2}^{(m)} \right]_{k,k}^* e^{-j\frac{ 2\pi \left( k-1 \right) \left( d_{2B}-d_{A2} \right)}{N}} \end{array}\right]. 
\end{equation}

\section{Estimation of the self-interference term in the JBD-DSTC scheme}  \label{sec:Diff_appendixC} 
As a first step we investigate the expected value of $\mbox{$ {D}_{B,k}^{(m)} $}^H {\bm Y}_{B,k}^{(m)}$ over the constellation points of ${S}_{A,k}^{(m)}$ and ${S}_{B,k}^{(m)}$. We can write this as $\E \left[ \mbox{$ {D}_{B,k}^{(m)} $}^H {\bm Y}_{B,k}^{(m)} \right] =  \E \left[ \mbox{$ {D}_{B,k}^{(m)} $}^H {D}_{B,k}^{(m)}\right] {\bm \mu}_{B,k}+
 \E \left[ \mbox{$ {D}_{B,k}^{(m)} $}^H {D}_{A,k}^{(m)} \right] {\bm \mu}_{A,k} + {\bm V}_{B,k}^{(m)}$.

To simplify exposition, and since we aim to take the expectation over the constellation points rather than time or frequency, we will drop the subcarrier index ($k$) and the block index ($m$) such that ${D}_{i,k}^{(m)}$, $\widetilde{\bm S}_{i,k,r}^{(m)}$ and $\widetilde{S}_{i,k,r}^{(m,t)} $ will be expressed by ${D}_{i}$, $\widetilde{\bm S}_{i,r}$ and $\widetilde{S}_{i,r}^{(t)}$, respectively.
We can write $\mbox{$ {D}_{B}^{} $}^H  {D}_{B}^{}$ as
\begin{IEEEeqnarray}{rCl}
 \mbox{$ {D}_{B} $}^H  {D}_{B} &=&\left[\begin{array}{cccc}
  \widetilde{\bm S}_{B,1}^{H} O_1^H O_1\widetilde{\bm S}_{B,1}^{}  & \widetilde{\bm S}_{B,1}^{H}O_1^H O_2\widetilde{\bm S}_{B,2}^{}   &  \hdots & \widetilde{\bm S}_{B,1}^{H}O_1^H O_{N_R}\widetilde{\bm S}_{B,N_R}^{}  \\
 \widetilde{\bm S}_{B,2}^{H}O_2^H O_1\widetilde{\bm S}_{B,1}^{}  & \widetilde{\bm S}_{B,2}^{H}O_2^H O_2\widetilde{\bm S}_{B,2}^{}   & \hdots & \widetilde{\bm S}_{B,2}^{H}O_2^H O_{N_R}\widetilde{\bm S}_{B,N_R}^{}  \\
  \vdots &  \vdots &  \ddots & \vdots \\
  \widetilde{\bm S}_{B,N_R}^{H}O_{N_R}^H O_1\widetilde{\bm S}_{B,1}^{}  & \hdots   &  \hdots &  \widetilde{\bm S}_{B,N_R}^{H}O_{N_R}^H O_{N_R}\widetilde{\bm S}_{B,N_R}^{}
     \end{array} \right], \\
     &=&\left[\begin{array}{cccc}
    T & \widetilde{\bm S}_{B,1}^{H}O_1^H O_2\widetilde{\bm S}_{B,2}^{}   &  \hdots & \widetilde{\bm S}_{B,1}^{H}O_1^H O_{N_R}\widetilde{\bm S}_{B,N_R}^{}  \\
 \widetilde{\bm S}_{B,2}^{H}O_2^H O_1\widetilde{\bm S}_{B,1}^{}  & T   & \hdots & \widetilde{\bm S}_{B,2}^{H}O_2^H O_{N_R}\widetilde{\bm S}_{B,N_R}^{}  \\
  \vdots &  \vdots &  \ddots & \vdots \\
  \widetilde{\bm S}_{B,N_R}^{H}O_{N_R}^H O_1\widetilde{\bm S}_{B,1}^{}  & \hdots   &  \hdots & T
     \end{array} \right],
\label{eqn:DBDB}
 \end{IEEEeqnarray}
where we used the fact $O_r^H O_r=I_T$. Let $J^{i,j}=O_i^H O_j$ and let its element in the $(l,p)$ position be denoted by $J_{l,p}^{i,j}$. Recall that $J^{i,j}$, $i \neq j$, is a hollow matrix, i.e., $J_{l,l}^{i,j}=0$ $\forall l\in \{1,2,\ldots,T\}$.

Note that $  \widetilde{\bm S}_{B,i}^{H} J^{i,j} \widetilde{\bm S}_{B,j}^{}= \sum_{r=1}^{T}\mbox{$   \widetilde{ S}_{B,i}^{(r)}$}^*  \sum_{c=1}^{T}  J_{r,c}^{i,j} \widetilde{ S}_{B,j}^{(c)}  = \sum_{r=1}^{T} \sum_{c=1}^{T}  J_{r,c}^{i,j} \mbox{$   \widetilde{ S}_{B,i}^{(r)}$}^*  \widetilde{ S}_{B,j}^{(c)} $. Hence, we can write $\E \left[  \widetilde{\bm S}_{B,i}^{H} J^{i,j} \widetilde{\bm S}_{B,j}^{} \right] =  \sum_{r=1}^{T} \sum_{c=1}^{T}  J_{r,c}^{i,j} \E \left[  \mbox{$   \widetilde{ S}_{B,i}^{(r)}$}^*  \widetilde{ S}_{B,j}^{(c)} \right]$. Due to the differential encoding, both $\widetilde{ S}_{B,i}^{(r)}$ and $\widetilde{ S}_{B,j}^{(c)} $ are correlated since they both consist of differently-weighted linear combination of the same $T$ random variables, which on the other hand, are also correlated with each other due to the same reason. However, by examining their correlation coefficients, we have found that they are small enough to be neglected. Therefore, we approximate their correlation by zero, and hence $\E \left[  \widetilde{\bm S}_{B,i}^{H} J^{i,j} \widetilde{\bm S}_{B,j}^{} \right]\approx 0$, $i \neq j$, and $ \E \left[ \mbox{$ {D}_{B} $}^H  {D}_{B} \right] \approx T I_{N_R}$. Following the same rationale, we conclude that $ \E \left[ \mbox{$ {D}_{B} $}^H  {D}_{A} \right] \approx 0_{N_R\times N_R}$.
%

Finally, assuming large $M$, we use the law of large numbers to approximate the expected value of $ \mbox{$ {D}_{B,k}^{(m)} $}^H {\bm Y}_{B,k}^{(m)}$ by its time average, which can be calculated at user B, as $ \sum_{m=1}^{M}  \mbox{$ {D}_{B,k}^{(m)} $}^H {\bm Y}_{B,k}^{(m)} / M $, and hence we obtain $\widehat{{\bm \mu}}_{B,k}
  \approx  \sum_{m=1}^{M}  \mbox{$ {D}_{B,k}^{(m)} $}^H {\bm Y}_{B,k}^{(m)} / (M T)$ for large SNRs.

\bibliographystyle{ieeetran}
\singlespace
\vspace{-.05cm}
{\fontsize{10}{12}\selectfont

\begin{thebibliography}{10}
\providecommand{\url}[1]{#1}
\csname url@samestyle\endcsname
\providecommand{\newblock}{\relax}
\providecommand{\bibinfo}[2]{#2}
\providecommand{\BIBentrySTDinterwordspacing}{\spaceskip=0pt\relax}
\providecommand{\BIBentryALTinterwordstretchfactor}{4}
\providecommand{\BIBentryALTinterwordspacing}{\spaceskip=\fontdimen2\font plus
\BIBentryALTinterwordstretchfactor\fontdimen3\font minus
  \fontdimen4\font\relax}
\providecommand{\BIBforeignlanguage}[2]{{%
\expandafter\ifx\csname l@#1\endcsname\relax
\typeout{** WARNING: IEEEtran.bst: No hyphenation pattern has been}%
\typeout{** loaded for the language `#1'. Using the pattern for}%
\typeout{** the default language instead.}%
\else
\language=\csname l@#1\endcsname
\fi
#2}}
\providecommand{\BIBdecl}{\relax}
\BIBdecl

\bibitem{Salim_2015}
A.~Salim and T.~M. Duman, ``An asynchronous two-way relay system with full
  delay diversity in time-varying multipath environments,'' in \emph{IEEE
  International Conference on Computing, Networking and Communications (ICNC)},
  Feb. 2015, pp. 900--904.

\bibitem{Duman_2015}
------, ``A delay-tolerant asynchronous two-way-relay system over
  doubly-selective fading channels,'' \emph{IEEE Transactions on Wireless
  Communications}, vol.~14, no.~7, pp. 3850--3865, July 2015.

\bibitem{Song_2010}
L.~Song, Y.~Li, A.~Huang, B.~Jiao, and A.~Vasilakos, ``Differential modulation
  for bidirectional relaying with analog network coding,'' \emph{IEEE
  Transactions on Signal Processing}, vol.~58, no.~7, pp. 3933--3938, 2010.

\bibitem{Cui_2009}
T.~Cui, F.~Gao, and C.~Tellambura, ``Differential modulation for two-way
  wireless communications: a perspective of differential network coding at the
  physical layer,'' \emph{IEEE Transactions on Communications}, vol.~57,
  no.~10, pp. 2977--2987, 2009.

\bibitem{Guan_2011}
W.~Guan and K.~Liu, ``Performance analysis of two-way relaying with
  non-coherent differential modulation,'' \emph{IEEE Transactions on Wireless
  Communications}, vol.~10, no.~6, pp. 2004--2014, 2011.

\bibitem{Zhu_2012}
K.~Zhu and A.~Burr, ``A simple non-coherent physical-layer network coding for
  transmissions over two-way relay channels,'' in \emph{IEEE Global
  Communications Conference (GLOBECOM)}, 2012, pp. 2268--2273.

\bibitem{Debbah2010}
L.~Song, G.~Hong, B.~Jiao, and M.~Debbah, ``Joint relay selection and analog
  network coding using differential modulation in two-way relay channels,''
  \emph{{IEEE} Transactions on Vehicular Technology}, vol.~59, no.~6, pp.
  2932--2939, 2010.

\bibitem{utkovski_2009_distributed}
Z.~Utkovski, G.~Yammine, and J.~Lindner, ``A distributed differential
  space-time coding scheme for two-way wireless relay networks,'' in \emph{IEEE
  International Symposium on Information Theory}, 2009, pp. 779--783.

\bibitem{Huo_DDSTC_TWR_2012}
Q.~Huo, L.~Song, Y.~Li, and B.~Jiao, ``A distributed differential space-time
  coding scheme with analog network coding in two-way relay networks,''
  \emph{{IEEE} Transactions on Signal Processing}, vol.~60, no.~9, pp.
  4998--5004, Sept 2012.

\bibitem{alabed_2013_distributed}
S.~Alabed, M.~Pesavento, and A.~Klein, ``Distributed differential space-time
  coding for two-way relay networks using analog network coding,'' in \emph{the
  21st European Signal Processing Conference (EUSIPCO)}, 2013, pp. 1--5.

\bibitem{Zhuo_diff_asynch_TWR_2013}
Z.~Wu, L.~Liu, Y.~Jin, and L.~Song, ``Signal detection for differential
  bidirectional relaying with analog network coding under imperfect
  synchronisation,'' \emph{{IEEE} Communications Letters}, vol.~17, no.~6, pp.
  1132--1135, June 2013.

\bibitem{qian2014asynchronous}
M.~Qian, Y.~Jin, Z.~Wu, and T.~Wang, ``Asynchronous two-way relaying networks
  using distributed differential space-time coding,'' \emph{International
  Journal of Antennas and Propagation}, vol. 2015, Article ID 563737, 9 pages,
  2015.

\bibitem{Divsalar_1990}
D.~Divsalar and M.~K. Simon, ``Multiple-symbol differential detection of
  mpsk,'' \emph{IEEE Transactions on Communications}, vol.~38, no.~3, pp.
  300--308, Mar 1990.

\bibitem{Hasna_2003}
M.~Hasna and M.-S. Alouini, ``End-to-end performance of transmission systems
  with relays over {R}ayleigh-fading channels,'' \emph{IEEE Transactions on
  Wireless Communications}, vol.~2, no.~6, pp. 1126--1131, Nov 2003.

\bibitem{Hassibi_DSTC_2006}
Y.~Jing and B.~Hassibi, ``Distributed space-time coding in wireless relay
  networks,'' \emph{IEEE Transactions on Wireless Communications}, vol.~5,
  no.~12, pp. 3524--3536, December 2006.

\bibitem{Jafarkhani_DDSTC_2008}
Y.~Jing and H.~Jafarkhani, ``Distributed differential space-time coding for
  wireless relay networks,'' \emph{IEEE Transactions on Communications},
  vol.~56, no.~7, pp. 1092--1100, July 2008.

\bibitem{Tarokh_1999}
V.~Tarokh, H.~Jafarkhani, and A.~Calderbank, ``Space-time block codes from
  orthogonal designs,'' \emph{IEEE Transactions on Information Theory},
  vol.~45, no.~5, pp. 1456--1467, Jul 1999.

\end{thebibliography}


}

\end{document}